\documentclass[pdflatex,sn-mathphys-num]{sn-jnl}

\usepackage{graphicx}
\usepackage{multirow}
\usepackage{amsmath,amssymb,amsfonts}
\usepackage{amsthm}
\usepackage{mathrsfs}
\usepackage[title]{appendix}
\usepackage{xcolor}
\usepackage{textcomp}
\usepackage{manyfoot}
\usepackage{booktabs}
\usepackage{algorithm}
\usepackage{algorithmicx}
\usepackage{algpseudocode}
\usepackage{listings}
\usepackage{tikz}
\usetikzlibrary{positioning,arrows.meta,calc,shapes.geometric,fit,backgrounds}
\usepackage{quantikz}
\usepackage{subcaption}
\usepackage{placeins}
\graphicspath{{fig/}{./}}

\theoremstyle{thmstyleone}
\newtheorem{theorem}{Theorem}
\newtheorem{proposition}[theorem]{Proposition}
\newtheorem{lemma}[theorem]{Lemma}
\newtheorem{corollary}[theorem]{Corollary}
\theoremstyle{thmstyletwo}

\newtheorem{remark}{Remark}
\theoremstyle{thmstylethree}
\newtheorem{definition}{Definition}

\tikzset{
  wC/.style={draw, thick, fill=gray!12, trapezium, trapezium angle=70, minimum height=0.55cm, font=\small, inner sep=1pt},
  wO/.style={wC, fill=red!12},
  wN/.style={wC, fill=blue!12},
  ty/.style={font=\scriptsize, inner sep=1pt},
  cup/.style={thick},
  outw/.style={very thick, red!70!black},
  every picture/.append style={>=Stealth}
}
\newcommand{\Pt}{\tilde{P}}
\newcommand{\Pp}{P_{\mathrm{post}}}
\newcommand{\smi}[1]{\texttt{#1}}
\newcommand{\Hthree}{H3}
\newcommand{\Height}{H8}

\begin{document}

\title[One molecule, many outputs; many molecules, one output]{One molecule, many outputs; many molecules, one output: exact diagnostics for atom-level compositional quantum encoders}

%% ---- Authors ----
\author*[1]{\fnm{Gautami Sanjay} \sur{Naik}}\email{gautaminaik2000@gmail.com}
\author[1]{\fnm{Ashwanth} \sur{Krishnan}}\email{ashwanth.krishnan@qpiai.tech}
\affil*[1]{\orgname{QpiAI India Pvt Ltd.}, \orgaddress{\city{Bengaluru}, \country{India}}}

\abstract{Quantum natural language processing (QNLP) turns the grammar of a sentence into a tensor network and then into a quantum circuit with trainable angles. We build such a model for molecules: each heavy atom is a word with one wire per bond, and each bond is a ``cup'' joining two wires, so the circuit's wiring is the molecular graph. We work out exactly what these circuits keep. With the typed angle sharing of lambeq, the circuits separate every isomer pair we tested. Yet for isomers that simpler sharing confuses, the output states stay almost the same, before and after training. On 7 Tox21 pairs and 1,988 MOSES pairs, nearly all the difference lies in how often the measurements succeed, which a standard readout ignores. Simpler sharing, by element and bond type, shows why: the output then depends only on a graph whose vertices are the bonds. This explains all but one of the identical-output pairs we certify, including 1.19\% of 232{,}191 MOSES molecules, and for alkyl side chains it reduces to a tree shape that tells chains apart up to 15 atoms but not from 16. On small pairs, this circuit and the 1-WL graph test each separate pairs the other misses. We also show that SMILES writing order can change the output, that a run succeeds with probability at most $2^{-n}$ for $n$ bonds, and that a cup-free design avoids this cost. No circuit beats a simple atom-count model on Tox21, and we claim no quantum advantage.}


\maketitle

%% =====================================================================
\section{Introduction}\label{sec:intro}

Quantum machine learning (QML) often uses parameterised quantum circuits, whose gates contain angles that are learned from data. Much of the field studies what such circuits can represent and how well they train \cite{biamonte2017qml,benedetti2019pqc,cerezo2021vqa,schuld2019feature,havlicek2019,mcclean2018barren}. Most designs use a fixed circuit template. A less common idea is to build the circuit from the structure of each input. The DisCoCat framework does this for sentences \cite{coecke2010}. A pregroup grammar \cite{lambek1999} gives every word a type, the way the types cancel in a sentence decides which tensors are joined, and toolkits such as lambeq turn this pattern into a circuit \cite{kartsaklis2021lambeq,lorenz2023qnlp,meichanetzidis2023grammar}.

Molecules look like a good fit for this idea. A SMILES string \cite{weininger1988} writes a molecule as a line of text, and the molecular graph behind it already says which atoms are bonded. The first use of QNLP in chemistry that we know of, by Kang and Kim \cite{kang2025mof}, took a different route. They wrote each metal--organic framework (MOF) as a three-word ``sentence'' (topology, metal node and organic linker) and found that a bag-of-words model, which ignores word order, did best among their QNLP models. They explained this by noting that the grammar follows word order, while the properties of a MOF depend on which parts are present. Studies on text have also found that models which ignore word order can do as well as grammar-based ones \cite{lorenz2023qnlp}.

In the MOF case the grammar carried no chemistry, so its weak result tells us little about a grammar built from chemical structure. Here we build such a grammar for small organic molecules and work out, as exactly as we can, what the resulting circuits compute. Our main question is not whether they predict well (on our data they do not). It is whether the output depends only on the molecule, and if not, where information is added or lost.

The construction is simple (Section~\ref{sec:construction}). Each heavy atom becomes a word whose wires are its bonds. Each bond becomes a cup that joins two of these wires. The one wire left open is the output qubit. Two questions guide the paper. If the same molecule is written as two different SMILES strings, do we get the same output? And can two different molecules give exactly the same output? For the simplest design the answers are ``no'' and ``yes''. A third question follows: once a fix makes two outputs different in principle, how many runs of the circuit are needed to tell them apart in practice?

We answer these questions with exact arithmetic and checked certificates rather than estimates. Our three main results are as follows.

\begin{figure}[t]
\centering
\includegraphics[width=\textwidth]{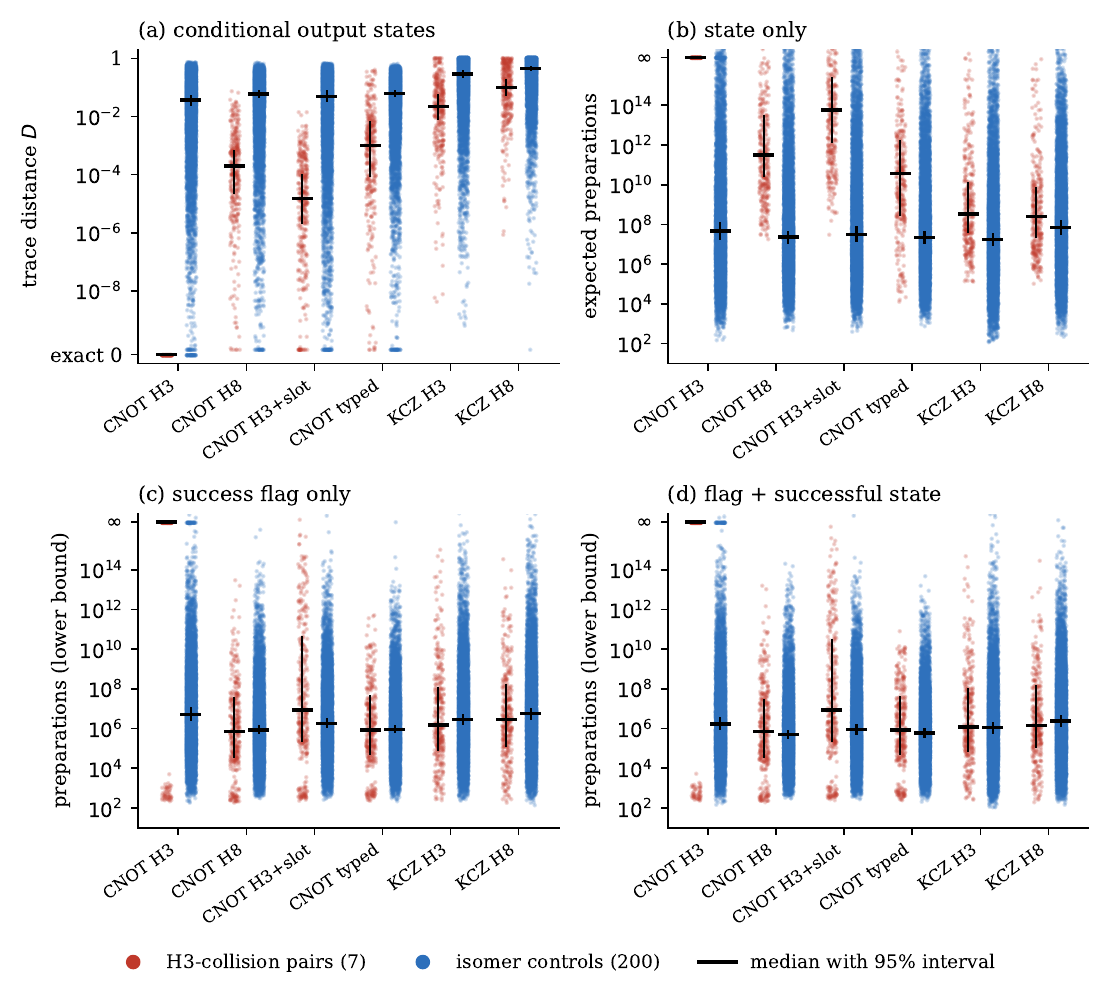}
\caption{Exact versus practical separation at 48 untrained tables (exact arithmetic). Red: the seven Tox21 pairs that collide under CNOT-\Hthree; blue: 200 control pairs. ``CNOT typed'' is our own typed scheme. (a) Trace distance between normalised output states, with exact zeros shown at the bottom. (b) Expected number of runs for the state-only protocol at 1\% error. (c) Lower bound when only the success flag is recorded. (d) Lower bound when the flag and the output state are both recorded. Black bars: medians with bootstrap 95\% intervals.}\label{fig:distinguish}
\end{figure}

\begin{enumerate}
\item \textbf{Fixed isomers differ in how often the circuit succeeds, not in what it outputs.} Each cup needs a measurement that succeeds only part of the time, and runs that fail are thrown away (this is called postselection). With typed sharing of angles, as in lambeq, which is what most users would choose, the circuits tell apart every pair of isomers we tested. But for the pairs that simpler sharing confuses, the output states are still almost the same. Their trace distance, a standard measure of how different two quantum states are, has a median of only $10^{-5}$--$10^{-3}$ on Tox21 and $10^{-9}$--$10^{-8}$ on 1,988 MOSES pairs, so telling them apart from the output qubit would take $10^{10}$ or more runs. Nearly all of the difference is in the success rate of the measurements, which a standard readout throws away (Fig.~\ref{fig:distinguish}, Table~\ref{tab:moses}; Sections~\ref{sec:res-lambeq} and \ref{sec:res-budget}). Training does not change this: after training on Tox21 these pairs stay as close as before, while ordinary isomers move further apart (Table~\ref{tab:trained}).
\item \textbf{One graph explains why the simplest design fails.} The simplest design gives two ports the same angle whenever they have the same element and bond type. It is the molecular version of sharing angles by word, and it uses the fewest parameters (13 on our data). Under this design, a network of CNOT-ladder words computes a sum over bit patterns, known as a spin-model partition function, on a graph that we call the \emph{port-line graph}. Its vertices are the bonds. Isomers with the same port-line graph give the same output for every choice of angles. This explains all but one of the identical-output pairs we certified in real molecule sets, from the Tox21 butyl acrylates to the 1.19\% of 232,191 MOSES molecules that have such a partner. For alkyl side chains it reduces to a tree shape, which tells all chains apart up to 15 atoms but not from 16, and the smallest failing pair is one that our compiler actually produces (Fig.~\ref{fig:portlinemain}; Sections~\ref{sec:portline}, \ref{sec:shape} and \ref{sec:res-collisions}). It also shows which small changes to the sharing are enough to remove the problem.
\item \textbf{The circuit and the 1-WL test see different things, on a small scale.} The 1-WL test, also called colour refinement, is the usual yardstick for graph neural networks. With a fixed canonical SMILES, neither method separates all the pairs that the other does. 1-WL separates every identical-output pair we certified, while the circuits separate six small constructed pairs, of at most seven atoms, that 1-WL cannot. On the MOSES set, 1-WL already separates everything (Fig.~\ref{fig:wl}; Section~\ref{sec:res-collisions}).
\end{enumerate}

Some further results are less surprising but complete the picture. The circuit wiring always matches the molecular graph; we check this against RDKit for every string. Even so, a CNOT word reads its wires in order, so the way a SMILES string is written changes the trained output. A symmetric word made of CZ gates removes this, at a cost in norm and gates (Fig.~\ref{fig:writing}; Section~\ref{sec:res-writing}). When the atom words are prepared separately, all $n$ bond measurements succeed together with probability at most $2^{-n}$ (Section~\ref{sec:bound}). A design with gates on the bonds instead of cups avoids this cost. If it reads only one atom it sees only the atoms nearby, but reading all atoms removes this limit (Section~\ref{sec:res-cupfree}). Finally, on the Tox21 NR-ER task no circuit beats a logistic regression on a simple count of atom types (Section~\ref{sec:res-prediction}).

Each claim carries a label that says what kind of evidence supports it: \textbf{[M]} a mathematical proof under stated conditions, \textbf{[F]} an exact result on a named finite set, \textbf{[N]} a numerical measurement, and \textbf{[X]} exploratory prediction results. All statements refer to the compiler, word types, sharing schemes and molecule sets defined below; we say this once here instead of repeating it. We do not claim any predictive or quantum advantage, and, following Bowles et al.\ \cite{bowles2024benchmark}, we compare every circuit with a simple classical baseline.
%% =====================================================================
\section{Background and related work}\label{sec:background}

\paragraph{Four related areas.} Four areas of work are often mixed up. \emph{Quantum chemistry} uses quantum computers to compute the electronic structure of molecules. \emph{Quantum machine learning} uses parameterised circuits as learning models. \emph{QNLP} builds those circuits from the grammar of a sentence. \emph{Quantum graph learning} lets a graph decide the layout of a circuit. This paper is QNLP-style quantum machine learning on molecular graphs; it does no electronic-structure calculation.

\paragraph{Pregroups and DisCoCat.} In a pregroup grammar, each word has a type made of basic types and their \emph{adjoints}. A noun has type $n$, and a transitive verb such as ``loves'' has type $n^{r}\, s\, n^{l}$, which says that it needs a noun on each side. A sentence is grammatical when neighbouring pairs such as $n\,n^{r}$ cancel until only the sentence type $s$ is left \cite{lambek1999}. DisCoCat turns each word into a tensor with one leg per basic type, and each cancellation into a \emph{cup} that joins two legs and sums over them \cite{coecke2010,coecke2017picturing}. On a quantum computer, a cup is a Bell measurement in which only one outcome, $\lvert\Phi^{+}\rangle$, is kept, so every cup lowers the chance that a run succeeds \cite{kartsaklis2021lambeq,lorenz2023qnlp}.

\paragraph{QNLP outside language.} Apart from the MOF study \cite{kang2025mof}, tensor-network circuits in the QNLP style have been used for protein sequences \cite{kundu2024protein}. We know of no earlier model that gives a type to each SMILES atom and joins atoms with one cup per bond, so we describe ours in full in Section~\ref{sec:construction}.

\paragraph{Circuits shaped by molecular graphs.} Several quantum models already use the bond graph to lay out a circuit. QMSE gives each atom a qubit and relies on canonical SMILES, because its output changes when the atoms are reordered \cite{boy2025qmse}. Pegg et al.\ put a CZ gate with symmetric rotations on every bond and share parameters by bond type \cite{pegg2026topology}. In our model a bond is instead a Bell measurement between the wires of two atom words, and the output is a single qubit on one chosen atom, the \emph{anchor}. Designs with gates on the bonds need no postselection; Section~\ref{sec:res-cupfree} tests a small one on the same molecule pairs.

\paragraph{Symmetry in quantum models.} Building symmetry into a model, so that its output does not change when the input is relabelled, is now a central design idea in quantum machine learning \cite{larocca2022group,meyer2023symmetry,nguyen2024equivariant}. Equivariant quantum graph circuits get this property from edge gates that are symmetric and commute with each other \cite{mernyei2022eqgc}, and related work makes gates symmetric by sharing their parameters \cite{mansky2025perm}. Our symmetric word (Section~\ref{sec:words}) uses the same idea inside each atom word.

\paragraph{Comparison with Weisfeiler--Leman.} Classical graph neural networks are usually compared with colour refinement, also called the 1-WL test, which repeatedly relabels each atom by the labels of its neighbours \cite{weisfeiler1968,xu2019gin,morris2019wl}. Quantum graph networks have recently been studied in the same way \cite{raj2026wl}. We use 1-WL as a reference for which molecule pairs a circuit with shared angles can or cannot tell apart.

\paragraph{Grammars in cheminformatics.} In cheminformatics, a ``molecular grammar'' usually means a set of rules for generating molecules that are guaranteed to be valid. Examples are a context-free grammar over SMILES \cite{kusner2017gvae}, the rules of SELFIES \cite{krenn2020selfies}, graph grammars learned from data \cite{guo2022graphgrammar}, and fragment vocabularies with typed attachment points \cite{jin2020hiervae,maziarz2022moler,kong2022psvae}. These grammars limit what can be generated. A DisCoCat grammar has a different job: it decides how the meaning is computed.

\paragraph{Three kinds of grammar.}\label{sec:three} So three kinds of rules are involved when a grammar is applied to molecules (Table~\ref{tab:three}), and much confusion comes from treating them as one. A \emph{representation} grammar decides which strings are well-formed. A \emph{chemical} grammar decides which combinations of atoms and bonds are valid. A \emph{compositional} grammar decides which tensor legs are joined.

\begin{table}[h]
\caption{Three kinds of grammar involved in compositional molecular models.}\label{tab:three}
\small\begin{tabular*}{\textwidth}{@{\extracolsep\fill}p{2.2cm}p{3.3cm}p{3.2cm}p{2.7cm}}
\toprule
Grammar & What its rules decide & Example rule & Enforced here by \\
\midrule
Representation & Which strings are well-formed SMILES or SELFIES & A ring-closure digit appears exactly twice & Our tokeniser and parser \\
Chemical & Which atom--bond combinations are chemically valid & An aliphatic carbon has bond-order sum at most 4 & RDKit sanitisation, plus an upper-bound check \\
Compositional (DisCoCat) & Which tensor legs are contracted & A port of type $b_1$ cancels with $b_1^{r}$ & Pregroup reduction, then contraction \\
\botrule
\end{tabular*}
\end{table}

Our construction lines up the compositional grammar with the bonds: one cup per bond. It does not make the types express chemical validity, which RDKit still checks. Nor does it escape the representation grammar, because the order of the wires and the decision to accept a string are both read from the SMILES string. Much of the paper measures what these partial matches cost.

%% =====================================================================
\section{The grammar: from SMILES to circuits}\label{sec:construction}

\subsection{Overview}\label{sec:pipeline}

\begin{figure}[t]
\centering
\includegraphics[width=\textwidth]{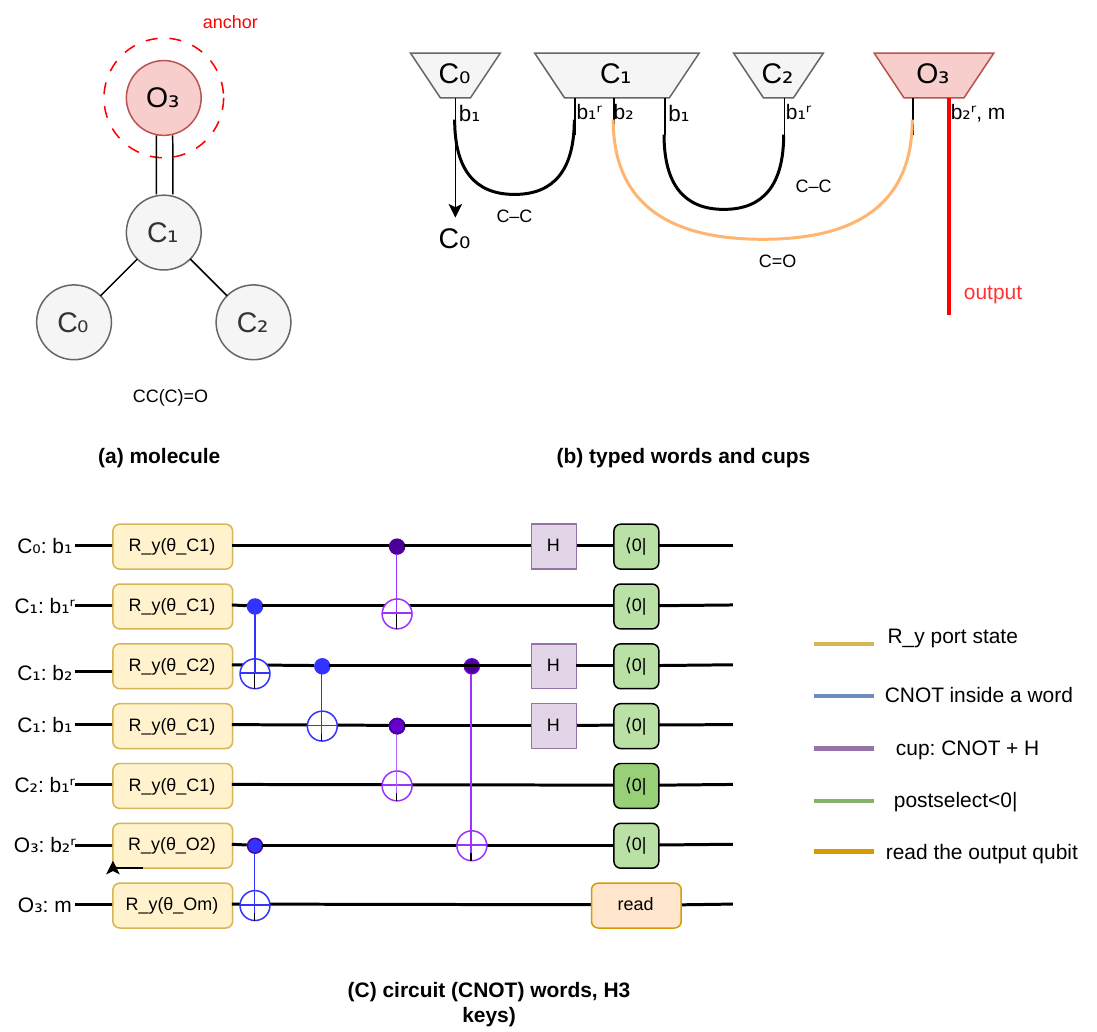}
\caption{From a molecule to a circuit. (a) Acetone, written \smi{CC(C)=O}. (b) The string diagram. Each heavy atom is a word (trapezium) with one wire per bond, labelled by bond type ($b_1$ single, $b_2$ double). The atom written second in each bond carries $b^{r}$, and each pair $b\,b^{r}$ is closed by a cup. The anchor O$_3$ also carries the output type $m$, the only open wire. (c) The circuit. Each wire is a qubit prepared by $R_y(\theta_k)$, where $k$ is the port's key. Each word applies a chain of CNOT gates over its ports, and each cup is a CNOT and a Hadamard followed by keeping only the outcome $\langle 00|$. Ports with the same key share one angle in all molecules.}\label{fig:overview}
\end{figure}

This section defines the grammar in full: which strings it accepts (Section~\ref{sec:scope}), which type each atom gets (Section~\ref{sec:lexicon}), which cancellations are allowed (Section~\ref{sec:rules}), and how the resulting diagram becomes a circuit (Section~\ref{sec:words}). Algorithm~\ref{alg:compile} summarises the compiler, and Fig.~\ref{fig:overview} shows its output for acetone. The types fix the wiring and the parameter key of every wire, which we call a \emph{port}. The gates inside each word are a separate modelling choice.

\begin{algorithm}[h]
\caption{Compiling a SMILES string to an anchored cup network}\label{alg:compile}
\begin{algorithmic}[1]
\Require SMILES string $s$ in the supported fragment (Section~\ref{sec:scope})
\State Tokenise and parse $s$; record, for each atom, its incoming tree bond, ring closures, ring openings and children
\State Sanitise with RDKit; check that parser atoms and RDKit atoms correspond index by index; read atom labels and bond classes
\State For each heavy atom, create a word whose ports are its bonds, in the order: incoming bond, closures, openings, children (reverse parse order)
\State Choose the anchor: the atom in the lowest-ranked singleton class of RDKit's symmetry ranks; reject $s$ if no class is a singleton
\State Append the output type $m$ to the anchor's word
\State Route: join the two ports of each bond with a cup, inserting swaps where needed, until only $m$ is open
\State \textbf{Guard:} execute the diagram and require the set of cup endpoints to equal the RDKit bond set
\State Prepare each port with key $k$ as $R_y(\theta_k)\lvert 0\rangle$, apply the word's gate family, and realise each cup as a Bell effect
\end{algorithmic}
\end{algorithm}

\subsection{The language: which strings are accepted}\label{sec:scope}

The grammar reads the atoms C, N, O, c, n, o, s and [nH]; single, double, triple and aromatic bonds, whether written or implied; and single-digit ring labels. A simple check rejects any non-aromatic C, N or O whose written bonds add up to more than 4, 3 or 2. This is only an upper limit, and RDKit decides whether the molecule is chemically valid. A string with lowercase (aromatic) atoms is accepted only if every ring system in it is a single, fully aromatic ring and no two ring systems are bonded directly. Charges, isotopes, stereochemistry, molecules in several pieces and fused aromatic rings are not covered.

This scope accepts about a quarter of Tox21 (Section~\ref{sec:data}); earlier, narrower versions of the grammar are described in Supplementary Note~S3. Because the check is made on the written string, acceptance can change when a molecule is written differently. For example, the Kekul\'e form of naphthalene is accepted, but its aromatic form is not. Of the 8,014 Tox21 rows, 2,356 are accepted as written, and rewriting every readable row in Kekul\'e form would add 833 and remove 18. All statements below about the same molecule written in different ways therefore refer to accepted strings.

\subsection{Basic types and the lexicon}\label{sec:lexicon}

Unlike a grammar for English, ours has no fixed dictionary. An atom's type is worked out from its place in the molecule, so the same element can get different types in different places. There are five basic types,
\begin{equation}
\mathcal{B} = \{\, b_1,\ b_2,\ b_3,\ b_a,\ m \,\},
\end{equation}
one for each bond type (single, double, triple, aromatic) and an output type $m$, which plays the role of the sentence type. Each $b_c$ has a partner $b_c^{r}$ (its right adjoint), and the only cancellation is $b_c\,b_c^{r}\to 1$.

\begin{definition}[Type assignment]\label{def:types}
Let $s$ be an accepted string with heavy atoms $a_0,\dots,a_{V-1}$ in the order they are written. After RDKit sanitisation, each bond $e=\{a_i,a_j\}$ has a class $c(e)\in\{1,2,3,a\}$. The atom $a_i$ receives one \emph{port} for each incident bond and, if it is the anchor, one extra port of type $m$. A port on bond $e$ has type
\begin{equation}
\tau_i(e) = \begin{cases} b_{c(e)} & \text{if the other atom of } e \text{ is written after } a_i \text{ (a \emph{plug})},\\[2pt] b_{c(e)}^{r} & \text{if it is written before } a_i \text{ (a \emph{socket})}. \end{cases}
\end{equation}
The ports are listed in a fixed order: the bond to the atom's parent in the SMILES tree, then ring closures, then ring openings, then child bonds in reverse order of writing, and finally $m$. The type of the word for $a_i$ is the product of its port types in this order.
\end{definition}

So every bond gives one plug, on the atom written first, and one socket, on the atom written second (Table~\ref{tab:lexicon}). Bond types are taken from RDKit after it has checked the molecule, so a Kekul\'e ring that RDKit treats as aromatic gets $b_a$ types.

\begin{table}[t]
\caption{Types given by Definition~\ref{def:types} for six molecules, one word per heavy atom in the order written. $\star$ marks the anchor, whose word ends in the output type $m$. The last column is the number of swaps needed before all types cancel.}\label{tab:lexicon}
\small\begin{tabular*}{\textwidth}{@{\extracolsep\fill}llp{7.0cm}c}
\toprule
Molecule & SMILES & Word types in written order & Swaps \\
\midrule
Ethanol & \smi{CCO} & C$^\star$: $b_1\,m$ \quad C: $b_1^{r}\,b_1$ \quad O: $b_1^{r}$ & 1 \\
Acetonitrile & \smi{CC\#N} & C$^\star$: $b_1\,m$ \quad C: $b_1^{r}\,b_3$ \quad N: $b_3^{r}$ & 1 \\
Acetone & \smi{CC(=O)C} & C: $b_1$ \quad C: $b_1^{r}\,b_1\,b_2$ \quad O$^\star$: $b_2^{r}\,m$ \quad C: $b_1^{r}$ & 1 \\
Acetone & \smi{CC(C)=O} & C: $b_1$ \quad C: $b_1^{r}\,b_2\,b_1$ \quad C: $b_1^{r}$ \quad O$^\star$: $b_2^{r}\,m$ & 0 \\
Oxirane & \smi{C1CO1} & C: $b_1\,b_1$ \quad C: $b_1^{r}\,b_1$ \quad O$^\star$: $b_1^{r}\,b_1^{r}\,m$ & 0 \\
Pyridine & \smi{c1ccncc1} & c$^\star$: $b_a\,b_a\,m$ \quad c: $b_a^{r}\,b_a$ \quad c: $b_a^{r}\,b_a$ \quad n: $b_a^{r}\,b_a$ \quad c: $b_a^{r}\,b_a$ \quad c: $b_a^{r}\,b_a^{r}$ & 2 \\
\botrule
\end{tabular*}
\end{table}

\subsection{Grammar rules and reduction}\label{sec:rules}

The whole grammar has seven rules. R1--R4 fix the words and their types, and R5--R7 say how the types cancel and when a string is accepted.

\begin{itemize}
\item[\textbf{R1}] \textbf{Words.} Every heavy atom is a word. Hydrogens are implicit and are not words.
\item[\textbf{R2}] \textbf{Ports.} A word has one port per incident bond, typed as in Definition~\ref{def:types}.
\item[\textbf{R3}] \textbf{Port order.} Parent bond, ring closures, ring openings, children in reverse writing order, then $m$.
\item[\textbf{R4}] \textbf{Output.} The anchor is the atom in the lowest-ranked singleton class of RDKit's symmetry ranks (\texttt{CanonicalRankAtoms} without tie-breaking); its word carries $m$. A molecule with no singleton class, such as cyclopropane, is rejected.
\item[\textbf{R5}] \textbf{Contraction.} $b_c\,b_c^{r}\to 1$, applied only to the plug and socket \emph{of the same bond}. Each contraction is drawn as a cup.
\item[\textbf{R6}] \textbf{Exchange.} Two adjacent types may swap places. Each swap is drawn as a wire crossing.
\item[\textbf{R7}] \textbf{Acceptance.} A string is accepted if it lies in the language of Section~\ref{sec:scope}, has an anchor, reduces to $m$ under R5--R6, and the set of cups equals the RDKit bond set (the guard in Algorithm~\ref{alg:compile}).
\end{itemize}

Two features differ from a textbook pregroup grammar. R6 lets wires cross, which is allowed in the mathematical setting where DisCoCat is usually defined \cite{coecke2010,coecke2017picturing}. And R5 matches ports by the bond they belong to, not by type alone. A rule based on type alone can join the wrong ports: in \smi{CC(=O)C}, a later $b_1^{r}$ could cancel against either of two open $b_1$ types.

\begin{proposition}[Every accepted string reduces]\label{prop:reduces} \textbf{[M]}
Every string in the language of Section~\ref{sec:scope} with an anchor reduces to $m$ under R5--R6, and the resulting cups are in one-to-one correspondence with its bonds.
\end{proposition}

\begin{proof}
Read the concatenated types from left to right, keeping a stack of open types. A plug is pushed. A socket belongs to a bond whose plug sits on an earlier atom, so that plug is already on the stack; move it next to the socket by swaps (R6) and contract (R5). Each bond is contracted exactly once, when its socket is read, and nothing else is contracted. The type $m$ is never contracted, so it is the only type left.
\end{proof}

For example, \smi{CC(=O)C} reduces as
\begin{align*}
b_1\cdot b_1^{r}\,b_1\,b_2\cdot b_2^{r}\,m\cdot b_1^{r}
&\;\to\; b_1\,b_2\,b_2^{r}\,m\,b_1^{r} && \text{(cup C$_0$--C$_1$)}\\
&\;\to\; b_1\,m\,b_1^{r} && \text{(cup C$_1$=O$_2$)}\\
&\;\to\; b_1\,b_1^{r}\,m \;\to\; m && \text{(swap, then cup C$_1$--C$_3$)},
\end{align*}
while \smi{CC(C)=O} cancels with nested cups and no swap (Fig.~\ref{fig:writing}). The number of swaps for a socket is the number of types sitting above its plug on the stack, which gives the counts in Table~\ref{tab:lexicon}.

\subsection{From diagrams to circuits}\label{sec:words}

As in DisCoCat and lambeq \cite{coecke2010,kartsaklis2021lambeq}, a fixed set of translation rules turns the diagram into a circuit (Table~\ref{tab:functor}).

\begin{table}[t]
\caption{How each element of the grammar becomes part of a circuit. $k$ is a word's number of ports, $n$ the number of bonds and $V$ the number of heavy atoms.}\label{tab:functor}
\footnotesize\begin{tabular*}{\textwidth}{@{\extracolsep\fill}p{2.6cm}p{3.5cm}p{5.0cm}}
\toprule
Grammar element & Tensor network & Circuit \\
\midrule
Basic type $b_c$ or $m$ & index of dimension 2 & one qubit \\
Word with $k$ ports & tensor with $k$ legs & $R_y(\theta_{\mathrm{key}})$ on each port qubit, then the word's gates (CNOT ladder or KCZ) \\
Cup ($b_c\,b_c^{r}\to1$) & contraction of two legs & CNOT, Hadamard, postselect $\langle 00|$ (Bell effect) \\
Swap & exchange of two legs & relabelling of wires (no gate) \\
Output $m$ & open leg & measure the anchor's $m$ qubit in the $Z$ and $X$ bases \\
\midrule
Whole molecule & contraction to a 2-vector & $2n+1$ qubits, $2n$ of them postselected \\
\botrule
\end{tabular*}
\end{table}

A molecule with $n$ bonds and $V$ heavy atoms has $2n$ bond ports and one output port, so its circuit uses $2n+1$ qubits, one $R_y$ gate per port, and one CNOT and one Hadamard per cup. Inside the words, CNOT words use $2n+1-V$ CNOTs and KCZ words use $\sum_i k_i(k_i-1)/2$ CZs. Acetone ($n=3$, $V=4$) needs 7 qubits, 3 CNOTs inside words and 3 CNOTs for the cups (Fig.~\ref{fig:overview}c). We computed all outputs exactly, by contracting the tensors, rather than by simulating measurements, so no number in this paper has sampling noise. Section~\ref{sec:res-budget} estimates how many measurements a real experiment would need.

We study two kinds of word. Both start with one $R_y$ rotation per port and differ in what follows:

\begin{itemize}
\item \textbf{CNOT word:} a chain (``ladder'') of CNOTs from port 0 to the last port, so port $j$ controls port $j{+}1$. The result depends on the order of the ports.
\item \textbf{KCZ word:} a CZ on every pair of ports. CZ gates are symmetric and can be applied in any order, so the result does not depend on the order of the ports.
\end{itemize}

\begin{figure}[t]
\centering
\includegraphics[width=0.92\textwidth]{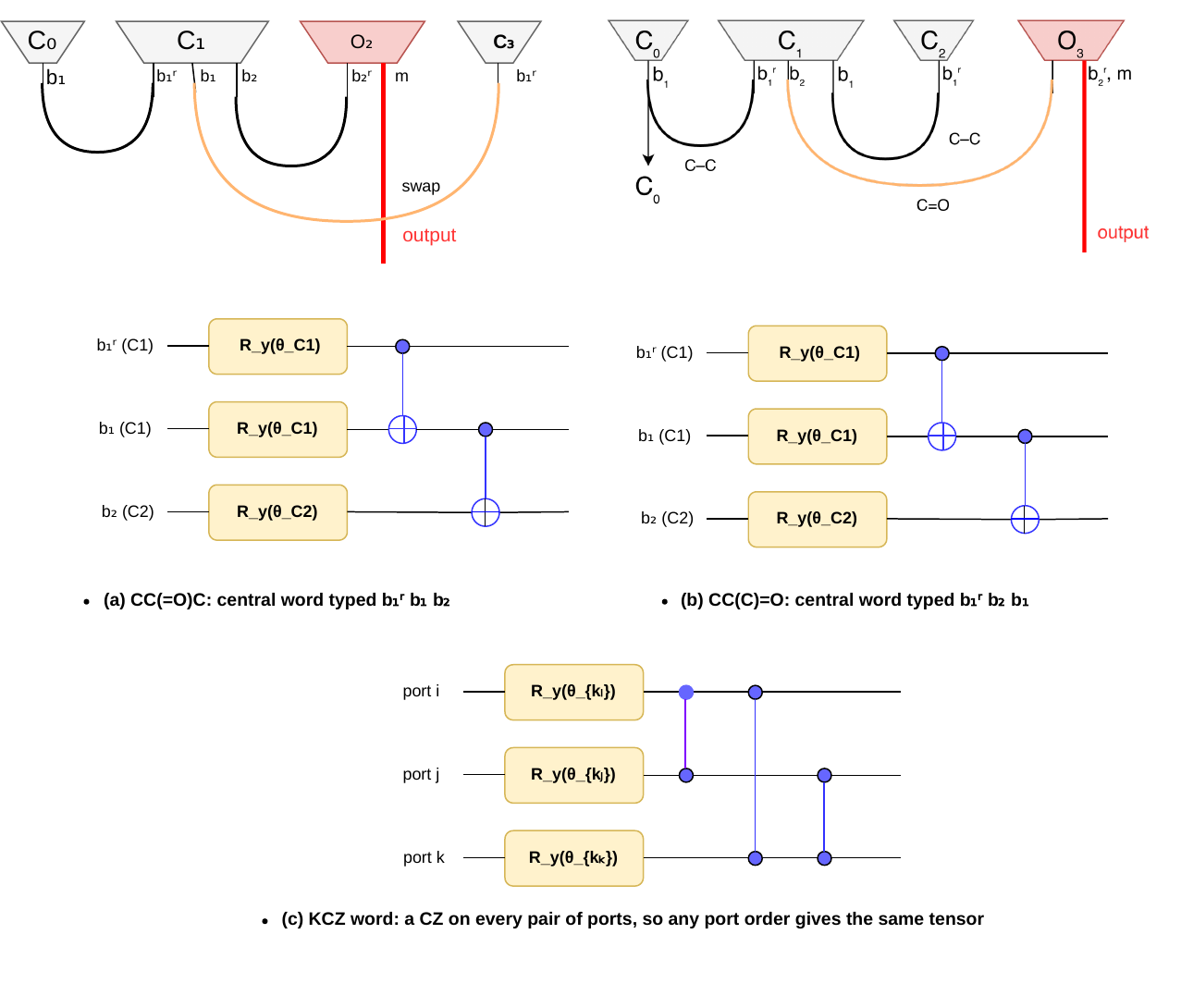}
\caption{Same molecule, different circuit. Acetone written in two ways gives the same cups but different string diagrams. In (a) the output wire crosses a cup, and the central word lists its single and double bonds in the opposite order, so its CNOT ladder (shown below each diagram) and the circuit output differ. (c) The KCZ word gives the same result for any order of its ports, which removes this dependence on writing (Proposition~\ref{prop:kcz}).}\label{fig:writing}
\end{figure}

A cup is a CNOT, a Hadamard and a measurement in which only the outcome $\lvert 00\rangle$ is kept; together these give the Bell effect $(\langle 00\rvert + \langle 11\rvert)/\sqrt{2}$. Fig.~\ref{fig:writing} shows why the choice of word matters. Writing acetone in another way leaves the cups unchanged but swaps the order of the central carbon's two child bonds. A CNOT ladder notices this; a KCZ word does not.

\subsection{Parameter sharing}\label{sec:keys}

Ports share angles through a \emph{key}. Under \Hthree{} the key is (atom label, bond type), with 40 possible keys. Under \Height{} it is (atom label, bond type, number of heavy-atom neighbours), with 240 keys. All ports with the same key, in every molecule, use the same angle. As a control, we also use an unshared model in which every port of every molecule has its own angle, and Section~\ref{sec:res-lambeq} adds five sharing schemes in the style of lambeq.

\subsection{Outputs, postselection and definedness}\label{sec:outputs}

Let $u_X(\theta)\in\mathbb{R}^{2}$ be the result of contracting molecule $X$'s network when each cup is replaced by a plain ``copy'' tensor (a delta tensor), and let $n_X$ be its number of bonds. Each Bell effect carries a factor $2^{-1/2}$, so the physical output amplitude is $v_X = 2^{-n_X/2}\,u_X$. We keep three quantities apart:
\begin{equation}
\Pt_X = \lVert u_X\rVert^{2}, \qquad \Pp(X) = \lVert v_X\rVert^{2} = 2^{-n_X}\,\Pt_X, \qquad X \text{ defined} \iff \Pt_X \ge 10^{-8}.
\end{equation}
$\Pp$ is the probability that all $n_X$ Bell measurements succeed. The ``defined'' threshold is only a bookkeeping rule. The readout uses the normalised state $v_X/\lVert v_X\rVert$, and a molecule whose output is undefined is given the average label of the training set; in the final \Height{} models this happened for 17 of the 1,319 development rows (Section~\ref{sec:res-prediction}).
%% =====================================================================
\section{What the circuits compute}\label{sec:theory}

This section works out what the circuits compute. Short proofs are given here; the longer ones are in Supplementary File~S2. We first fix two words. Two different molecules \emph{collide} if their outputs point in the same direction for every choice of angles, so that no readout of the output qubit can tell them apart. They are \emph{strongly equal} if their outputs are exactly the same.

\begin{definition}[Collision]
Molecules $X$ and $Y$ \emph{collide} under a word family and key scheme if $u_X(\theta)$ and $u_Y(\theta)$ are parallel for every $\theta$ and neither is identically zero. They are \emph{strongly equal} if $v_X \equiv v_Y$.
\end{definition}

A collision means the two molecules give the same normalised output state, so no readout of that state can tell them apart. How often the circuit succeeds is a separate matter: two outputs that point in the same direction but have different lengths succeed at different rates (Section~\ref{sec:res-budget}). Strong equality, which is what we certify, makes both the state and the success rate the same.

\begin{lemma}[Exact separation test]\label{lem:exact} \textbf{[M]}
Suppose every word is a layer of $R_y$ port states followed by a map with entries in $\{0,\pm 1\}$, and every cup is a delta tensor. Then each entry of $u_X$ is an integer polynomial in the half-angle cosines and sines. A single rational point on the parameter torus at which $\det[u_X,u_Y]\neq 0$ certifies that $X$ and $Y$ do not collide.
\end{lemma}

Choosing angles whose sines and cosines are fractions (from Pythagorean triples such as $3,4,5$) makes this test exact, with no rounding. Such a point, called a \emph{witness}, proves that two outputs differ somewhere. It does not say by how much.

\subsection{CNOT words compute partition functions on a port-line graph}\label{sec:portline}

Write $a_k(0)=\cos(\theta_k/2)$ and $a_k(1)=\sin(\theta_k/2)$ for the two amplitudes of the state $R_y(\theta_k)\lvert 0\rangle$.

\begin{definition}[Port-line graph]
For a compiled molecule, the \emph{port-line graph} $L$ has one vertex for each bond and one for the output $m$. Each word contributes a field on the vertex of its port-0 bond, labelled by that port's key, and an edge between the vertices of each pair of consecutive ports $(j-1,j)$, labelled by the key of port $j$. The vertex $m$ is the root.
\end{definition}

\begin{theorem}[Port-line partition function]\label{thm:portline} \textbf{[M]}
For any network of CNOT-ladder words joined by delta cups,
\begin{equation}\label{eq:portline}
u(y_m) \;=\; \sum_{y \in \{0,1\}^{\text{bonds}}} \;\prod_{\text{words } W} a_{k_{W,0}}\!\big(y_{W,0}\big) \prod_{j\ge 1} a_{k_{W,j}}\!\big(y_{W,j}\oplus y_{W,j-1}\big),
\end{equation}
where $y_{W,j}$ is the bit on the vertex of $L$ that carries port $j$ of word $W$. This is a two-state spin-model partition function on $L$ with fields and edge weights fixed by the keys.
\end{theorem}

\begin{proof}
A CNOT ladder maps a basis input $x$ to its prefix parities, $y_j = x_0\oplus\cdots\oplus x_j$ \cite{tserkis2025cnot}. This map is invertible with $x_0=y_0$ and $x_j = y_j\oplus y_{j-1}$. The input is the product state $\bigotimes_j R_y(\theta_{k_j})\lvert 0\rangle$, so the word's tensor is $T(y) = a_{k_0}(y_0)\prod_{j\ge1}a_{k_j}(y_j\oplus y_{j-1})$. A delta cup forces its two ports to carry the same bit, so each bond contributes a single summed variable. Multiplying the word tensors and summing the bond variables gives Eq.~\eqref{eq:portline}.
\end{proof}

\begin{corollary}\label{cor:portline} \textbf{[M]}
If two molecules have isomorphic labelled, rooted port-line graphs, their outputs are identical for every parameter value, so they are strongly equal.
\end{corollary}

\begin{figure}[t]
\centering
\includegraphics[width=0.9\textwidth]{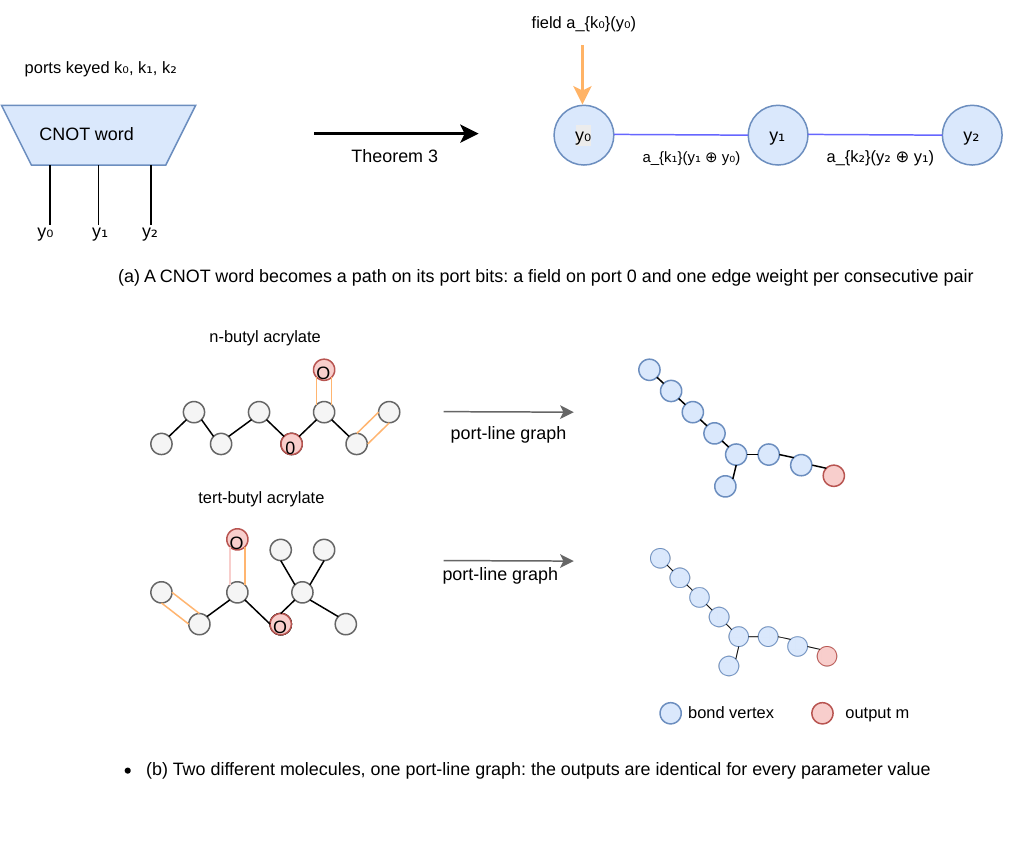}
\caption{Different molecules, same output. (a) The CNOT ladder turns a word's tensor into a product of terms, one for each pair of neighbouring ports. Gluing the words along their cups gives a spin model on the port-line graph, whose vertices are the bonds plus the output. (b) n-Butyl and tert-butyl acrylate under \Hthree{} keys (output vertex in red). Their port-line graphs are the same, so their outputs are the same (Corollary~\ref{cor:portline}) even though their circuits differ. Every carbon port has the same key, and the graph only records which bond follows which inside each word.}\label{fig:portlinemain}
\end{figure}

Results of this kind are known for related networks \cite{vandennest2007spin}; what is new here is the graph $L$ for this particular compiler. Under \Hthree, $L$ no longer records which atom a port belongs to when the atom labels and bond types are the same. This is how two different isomers can end up with the same $L$ (Fig.~\ref{fig:portlinemain}).

\subsection{What a shared-angle substituent retains}\label{sec:shape}

To see exactly what is lost, take an alkyl side chain (a substituent made only of carbon atoms joined by single bonds) under \Hthree. Here every port has the same key, and for each atom the bond pointing back towards the rest of the molecule is its port 0. A standard trick turns the ordered tree of the side chain into a binary tree: each atom's first child becomes its ``left child'' and its next sibling becomes its ``right sibling''. We call the result, with left and right forgotten, the \emph{shape} of the side chain.

\begin{theorem}[Shape factorisation] \textbf{[M]}
The substituent's contribution depends only on the unordered left-child/right-sibling (LCRS) shape of its ordered tree. If two molecules share an identical anchored remainder and their substituents have the same unordered LCRS shape, their outputs are equal. Substituents with $n$ atoms take at most $W(n)$ distinct values, where $W(n)$ are the Wedderburn--Etherington numbers. Consequently, under any writing policy of this form, at least 410 of the 507 rooted alkyl substituents with $n=10$ share their output with another.
\end{theorem}

In plain terms, the model cannot tell an atom's own branch apart from the other branches at the same parent. Is the shape the \emph{only} thing that survives? Not always:

\begin{itemize}
\item For shapes in which no node has two non-leaf children, the shape is complete for all $n$ \textbf{[M]}.
\item For all shapes with $n\le 15$, exhaustive exact checks find the shape complete, for example 451 of 451 classes at $n=12$ and 4,850 of 4,850 at $n=15$ \textbf{[F]}.
\item At $n=16$ exactly one pair of the 10,905 shapes has identical non-zero outputs, so 16 is the smallest size at which completeness fails; this pair and a pair at $n=17$ were each certified in four independent ways, and by gluing unary chains the shape is incomplete for every $n\ge 16$ \textbf{[M, F]}.
\end{itemize}

The failure has a simple cause. A carbon with a single child passes up a number that does not depend on where that child's sub-tree is rooted (it equals $t(1+t)$ times a sum over the whole sub-tree). Two sub-trees that differ only in their root can therefore be swapped, and when two such swaps balance each other at a branching carbon, two different shapes give the same output. So, unlike some other polynomials for rooted trees \cite{liu2019tree,burghart2024trees}, this one does not tell all trees apart.

\subsection{The symmetric word: invariance and cancellation}\label{sec:kcz}

\begin{proposition}[Writing invariance of KCZ] \textbf{[M]}\label{prop:kcz}
Permuting the ports of a KCZ word together with their keys leaves its tensor unchanged. Consequently, two accepted strings of the same molecule give identical KCZ outputs for every parameter table, provided (A1) keys are functions of graph attributes; (A2) the anchor rule is invariant under isomorphism; (A3) ports, bonds and cups correspond one-to-one and the compiled labelled graphs are isomorphic; (A4) cups are delta tensors and swaps contribute no tensor; and (A5) arithmetic is exact.
\end{proposition}

The proof matches up the two networks atom by atom, keeping the anchor fixed. CNOT words do not have this property, and the result says nothing about strings that the grammar rejects.

Symmetry has a price. The KCZ output is a sum of positive and negative terms, and these can cancel:

\begin{theorem}[KCZ as a signed Ising sum] \textbf{[M]}\label{thm:ising}
For all angles,
\begin{equation}
u(y) = f_m(y)\sum_{x}\prod_{e} g_e(x_e)\,(-1)^{Q_0(x) + y\,\ell(x)},
\end{equation}
where the sum runs over bond assignments, $Q_0$ is the adjacency quadratic form of the line graph $L(G)$ of the heavy-atom graph, $\ell$ marks the bonds at the anchor, and $g_e$ is the product of the two port amplitudes on bond $e$. At the balanced point $\theta=\pi/2$, $\Pt$ takes one of the values $2^{-r}$, $2^{-(r+1)}$ or $0$, where $r$ is the rank over $\mathrm{GF}(2)$ of the line-graph adjacency matrix; the radical and the anchor term decide which.
\end{theorem}

Each CZ adds a minus sign when both of its ports are 1, so a word whose ports hold $w$ ones picks up the sign $(-1)^{w(w-1)/2}$. Adding these signs over all words gives $Q_0$. This is a standard sum over bit patterns, and it explains the losses in norm reported in Section~\ref{sec:res-norm}.

\subsection{Postselection cost}\label{sec:bound}

Each cup multiplies the amplitude by $2^{-1/2}$, so $\Pp = 2^{-n}\Pt$. We fix how the circuit is run: (E1) each atom word is prepared separately, as a normalised state on its own port qubits; (E2) every bond is measured in the Bell basis, and a run is kept only if every bond gives $\lvert\Phi^{+}\rangle$; (E3) the output qubit of a kept run is read, and the whole procedure is repeated. Under (E1)--(E3), $\Pp$ is the chance that a run is kept. The bound below is about this procedure, not about every possible way of computing the same result \cite{arad2010tn}.

\begin{theorem}[Postselection bound] \textbf{[M]}\label{thm:bound}
Let the molecular graph have $n$ bonds. Under (E1)--(E3), for any unit-norm word states, any parameters, any sharing scheme and any ansatz,
\begin{equation}\label{eq:bound}
\Pt \le 1, \qquad\text{equivalently}\qquad \Pp \le 2^{-n}.
\end{equation}
The bound is attained on every graph, for example when every port is prepared in $\lvert 0\rangle$.
\end{theorem}

\begin{proof}
Contract the network step by step. At each step take two of the current tensors and contract, at the same time, every index they share. Writing $A$ with free indices $a$ and shared indices $s$, and $B$ with shared indices $s$ and free indices $b$, the result is $C_{ab}=\sum_s A_{as}B_{sb}$, a matrix product after reshaping, so $\lVert C\rVert_F\le\lVert A\rVert_F\,\lVert B\rVert_F$. Because every shared index is contracted when two pieces merge, no tensor ever has two of its own legs joined, and every step has this form. The words start with unit norm, so the final two-component vector obeys $\lVert u\rVert\le1$, that is $\Pt\le1$, and $\Pp=2^{-n}\Pt\le2^{-n}$. If every port is $\lvert 0\rangle$, each word is $\lvert0\cdots0\rangle$ for both word families and $u=(1,0)$, so $\Pt=1$.
\end{proof}

Since $n=V-1+\beta$ for a molecule with $\beta$ rings, the theorem is stronger than a bound based on the number of atoms, $\Pp\le2^{-(V-1)}$, by a factor $2^{-\beta}$: a bond that closes a ring costs as much as any other bond.

\begin{figure}[t]
\centering
\includegraphics[width=0.8\textwidth]{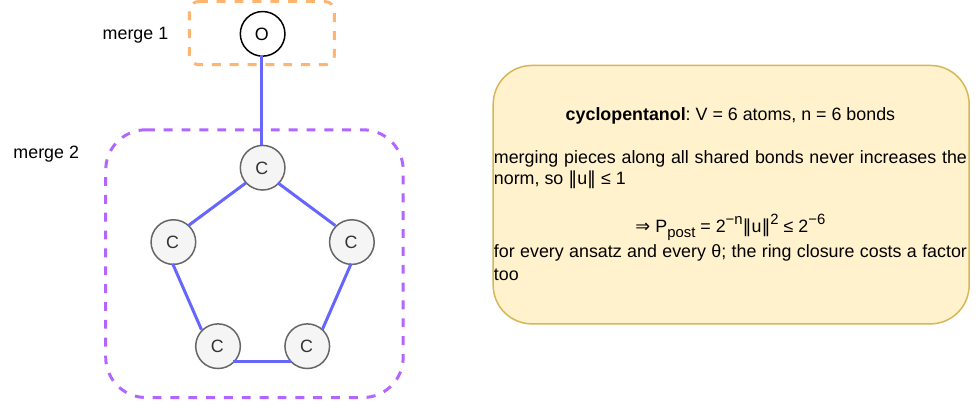}
\caption{The postselection bound (Theorem~\ref{thm:bound}). Each merge joins two pieces along all the bonds between them. It is a matrix product, so it cannot make the norm larger, and normalised words therefore give $\lVert u\rVert\le1$. Each of the $n$ Bell measurements, including those on ring-closing bonds, multiplies $\Pp$ by $\tfrac12$.}\label{fig:spanning}
\end{figure}

On average, then, at least $2^{n}$ runs are needed for each kept run: about $10^{6}$ for 20 bonds and $10^{9}$ for 30. The inequality itself follows from a standard property of matrix norms, and the link between evaluating tensor networks by projections and the norms met along the way goes back to Arad and Landau \cite{arad2010tn}; postselection is also known to be both powerful and costly in general \cite{aaronson2005postbqp}. What matters for circuits with one cup per bond is that no choice of gates, angles or sharing avoids the $2^{n}$ cost. Reaching the bound does not make the output useless: if every bond port stays in $\lvert0\rangle$ but the anchor's output port is rotated by an angle that depends on its label, the output still records that label.

\begin{remark}[Failed outcomes are not noise] \textbf{[M]}
For CNOT and KCZ words, a failed Bell outcome gives, up to a sign, the same network with some angles reflected ($\theta\to-\theta$, $\pi-\theta$ or $\pi+\theta$ on certain ports). This follows from the standard rules for moving Pauli corrections through such circuits \cite{gottesman1999teleport,raussendorf2003mbqc}. Failed runs therefore still carry information, but each one breaks the sharing of angles. Section~\ref{sec:res-budget} uses only whether a run succeeded or failed.
\end{remark}

\begin{remark}[Classical cost of exact evaluation] \textbf{[M, F]}
A tensor network can be contracted on a classical computer in time that grows exponentially only in a measure called the treewidth \cite{markov2008tn}, taken here on the line graph of the molecular graph with one extra vertex for the output. For all 232,191 molecules of the MOSES set (Section~\ref{sec:data}) this treewidth is at most 5, and at most 2 for 87\% of them. So, for these molecules, computing the output at given angles is cheap on an ordinary computer.
\end{remark}
%% =====================================================================
\section{Data and protocols}\label{sec:data}

We used five sets of molecules, each to answer a different question. Only Tox21 was ever used with labels (Table~\ref{tab:populations}).

\begin{table}[h]
\caption{Molecule populations used in this study.}\label{tab:populations}
\small\begin{tabular*}{\textwidth}{@{\extracolsep\fill}p{3.0cm}p{3.5cm}p{3.9cm}l}
\toprule
Population & Size & Purpose & Labels \\
\midrule
Pilot cohort & 34 molecules, 187 strings & Early traversal and collision checks & No \\
Tox21 NR-ER development & 1,319 rows; 1,299 identities; 12,614 strings & Training, writing dependence, first collision census & Yes \\
Generated alkyl set & 2,524 strings; 4 remainders; alkyl sizes 3--40 & Generality and mechanism of collisions & No \\
Engineered strings & 21 candidates & Compiler realisation of the $n=16$ case & No \\
MOSES external pool & 1,936,962 rows $\to$ 232,191 identities & External census, writing sensitivity, norms & No \\
\botrule
\end{tabular*}
\end{table}

\paragraph{Tox21.} We used the DeepChem copy of Tox21 \cite{wu2018moleculenet} (8,014 rows) and its NR-ER assay. We removed invalid rows, rows without an NR-ER label and duplicates with conflicting labels, which left 6,232 rows, and split these by scaffold. We set aside 934 rows, grouped by scaffold, and did not use them in this paper. Of the other 5,298 rows, 1,560 were accepted by the grammar and compiled: 1,319 development rows (181 positive), split into five folds so that each molecule appears in only one fold, and 241 validation rows (15 positive). For each development molecule we made up to eight other writings with RDKit random traversals and renumbered ring digits, 12,614 strings in total.

\paragraph{Generated set.} We joined four fixed parts (acrylate, methyl ester, nitrile and amide) to every allowed alkyl tree, written in every allowed way, up to $n=8$ (set S1, 1,884 strings), and to selected trees for $n=9$--$40$ (set S2, 640 strings). The collision census compares strings that share the same set, fixed part and alkyl size (215,924 pairs per arm). The typed-sharing test of Section~\ref{sec:res-lambeq} compares every pair of strings in S1 (1,773,786 pairs).

\paragraph{MOSES pool.} We passed all 1,936,962 MOSES rows \cite{polykovskiy2020moses} through the grammar. It accepted 232,191 distinct molecules, which gives about $2.7\times10^{10}$ pairs per arm. Only 290 of these molecules have no ring, so the alkyl side-chain effect of Section~\ref{sec:shape} hardly appears in this set.

\paragraph{Parameters.} Studies without labels used made-up angle tables: angles with rational sines and cosines for the exact tests, and, for the norm studies, angles drawn for each key either from the starting range $\theta\sim U(0.15\pi,0.85\pi)$ or from $U[0,2\pi)$, on a sample of 512 MOSES molecules chosen to cover their range of sizes. Each fold model was trained once, from a fixed starting table, with 2,000 Adam steps on the usual cross-entropy loss plus a penalty ($\lambda=0.01$) that pushes $\log_{10}\Pt$ to stay above $-6$, and batch size 16. A linear readout acts on $(\langle X\rangle,\langle Z\rangle)$ of the normalised output qubit. We used RDKit 2026.03.6 \cite{rdkit}, lambeq 0.5.0 \cite{kartsaklis2021lambeq} and Qiskit 2.5.2 \cite{qiskit2024}.

\paragraph{Certification.} Every ``separated'' verdict is backed by an exact witness (Lemma~\ref{lem:exact}). Every ``equal'' verdict is backed either by a symbolic identity between the two polynomials or by an explicit mapping between the two port-line graphs, and a separate program checks it. In the MOSES census, 1,410 of the 1,412 equal pairs were certified in both ways. Independently written code re-checked the certificates for every major set. Supplementary Note~S3 describes the census pipeline, the analysis protocols and the corrections we made along the way.

\paragraph{Use of AI tools.} Generative AI tools were used to help draft and edit the manuscript text and as coding assistants for the analysis and plotting scripts. The authors checked all code, results and text and take full responsibility for them.

%% =====================================================================
\section{Results}\label{sec:results}

\subsection{Compilation is graph-faithful}\label{sec:res-faithful}

Every accepted string (all 12,614 Tox21 development strings, 2,524 generated strings and 232,191 MOSES molecules) compiled into a network whose cups are exactly the bonds that RDKit finds \textbf{[F]}. Renumbering the ring-closure digits left every output unchanged, bit for bit.

\subsection{Outputs still depend on how the molecule is written}\label{sec:res-writing}

A CNOT word reads its ports in an order that depends on how the molecule is written. In trained models we measure this with the \emph{traversal variance fraction}, $\mathrm{TVF}=W/(W+B)$. Here $W$ is how much the model's score (logit) varies between writings of the same molecule and $B$ is how much it varies between molecules. A value of 0 means the writing never matters.

\begin{table}[h]
\caption{Writing dependence of CNOT and KCZ outputs.}\label{tab:writing}
\begin{tabular*}{\textwidth}{@{\extracolsep\fill}p{4.4cm}p{1.9cm}p{5.6cm}}
\toprule
Setting & Word & Result \\
\midrule
Trained, Tox21, \Height{} keys & CNOT & TVF fold mean 0.224 (SD 0.136; folds 0.063--0.432); median per-molecule logit range 0.105, max 3.79 \textbf{[N]} \\
Trained, Tox21, \Hthree{} keys & CNOT & TVF fold mean 0.270 (SD 0.209) \textbf{[N]} \\
MOSES, 1,000 molecules $\times$ 4 writings & CNOT (\Hthree, \Height) & 1,000 of 1,000 writing-sensitive, each proved by an exact non-zero determinant \textbf{[F]} \\
Same sample & KCZ (\Hthree, \Height) & 0 of 1,000 differ at 3 tables \textbf{[N]} \\
Generated set & KCZ & 2,852 within-molecule pairs certified strongly equal per key scheme \textbf{[F]} \\
Tox21, trained KCZ & KCZ & All 50,456 output comparisons agree (max relative deviation $7.2\times10^{-14}$) \textbf{[N]} \\
\botrule
\end{tabular*}
\end{table}

Keys that also record the number of neighbours do not remove this dependence on writing; the symmetric word does, as Proposition~\ref{prop:kcz} predicts (Table~\ref{tab:writing}). The TVF leaves out 65 of the 1,299 molecules, 50 of them because the output was defined for some writings and not for others, so if anything it understates the effect.

\subsection{Coarse sharing makes distinct molecules identical}\label{sec:res-collisions}

Under \Hthree{} keys, the CNOT model gives distinct molecules the same output, for every choice of angles, in every set we tested (Table~\ref{tab:collisions}). Under \Height{} keys, with KCZ words or without sharing, every pair in these sets is separated \textbf{[F]}. The one constructed exception for \Height{} is shown in the 1-WL comparison below.

\begin{table}[h]
\caption{Certified collisions per arm. Every pair not listed as colliding is separated by an exact witness; no pair was left unresolved.}\label{tab:collisions}
\footnotesize\begin{tabular*}{\textwidth}{@{\extracolsep\fill}lrrrrr}
\toprule
Population & Pairs per arm & CNOT-\Hthree & CNOT-\Height & KCZ (both) & Unshared \\
\midrule
Pilot (two isomers) & 64 & 7 & 0 & -- & 0 \\
Tox21 dev., canonical & 843,051 & 7 & 0 & 0 & 0 \\
Generated alkyl set & 215,924 & 8,671$^{a}$ & 0 & 0 & 0 \\
Engineered $n=16,17$ & 2 & 2 & 0 & 0 & 0 \\
MOSES external pool & $2.7\times10^{10}$ & 1,412 & 0 & 0 & -- \\
\botrule
\end{tabular*}
\footnotetext{$^{a}$ String pairs, covering 6,393 molecule pairs. A dash means the arm was not run.}
\end{table}

\paragraph{Tox21.} The seven colliding pairs involve eleven molecules (Appendix~\ref{app:tox21}), among them the three butyl acrylates and the 2,4- and 2,6-isomers of both diaminotoluene and toluene diisocyanate. Corollary~\ref{cor:portline} explains six of the pairs. The seventh, tetrahydrofuran and dibutyl ether, is equal only up to scale, and it cannot be strongly equal. That would need $u_X = 2^{(n_X-n_Y)/2}u_Y$, and with 5 and 8 bonds this factor is irrational, while at rational angles both outputs are rational. The two ethers therefore succeed at different rates (Section~\ref{sec:res-budget}); we do not have a derived explanation for this pair. For every pair there is a certified choice of angles at which both outputs are defined, so these collisions do reach real predictions. A purely numerical check would have missed two of the seven.

\paragraph{Generated set.} Section~\ref{sec:theory} explains all 8,671 certified collisions: 2,168 by a substitution rule for caterpillar-shaped chains, 2,797 by the shape result and 3,706 (where the anchor lies inside the side chain) by Corollary~\ref{cor:portline}. Among acrylates with 7- or 8-carbon alkyl groups, every molecule collides with at least one other, and 322 colliding pairs are between two RDKit canonical strings, so unusual writings are not needed to cause them.

\paragraph{Beyond the port-line mechanism.} The compiler does produce the $n=16$ case of Section~\ref{sec:shape}. The C$_{19}$H$_{36}$O$_2$ acrylates \smi{C=CC(=O)OC(CC(CCCCCC)CC)CC(C)CC} and \smi{C=CC(=O)OC(CCCC)CC(CCCC(C)CC)CC} give identical, non-zero polynomial outputs even though their port-line graphs are different, and the same holds for a C$_{20}$H$_{38}$O$_2$ pair at $n=17$. Their RDKit canonical strings are separated, so this collision depends on how the molecules are written.

\paragraph{External pool.} Across the 232,191 MOSES molecules, CNOT-\Hthree{} has 1,412 strongly equal pairs, covering 2,760 molecules (1.19\%). Every other pair is separated by an exact witness. All 1,412 pairs have matching port-line graphs. Typical cases are butyl and propyl isomers and ring position isomers such as 2,4- and 2,6-dimethylanilides. No collision of the $n=16$ kind occurs in this set.

\paragraph{Comparison with 1-WL.} Because the CNOT output depends on the string, we compare one fixed pipeline: molecule, RDKit canonical SMILES, compiler and anchor rule, and one shared angle table. 1-WL colours atoms by element and bonds by type; its anchored version also gives the anchor atom its own colour. We decide exactly whether two molecules are 1-WL equivalent by refining their colours together. Both the canonical writing and the anchor add information before any gate acts, so this comparison is about the whole pipeline. The KCZ arms, which do not depend on the writing, are the cleaner case for comparing architectures. We find three things \textbf{[F]}.

\begin{enumerate}
\item \emph{1-WL separates every collision found earlier.} All 10,092 certified CNOT-\Hthree{} collision records are separated by 1-WL, plain or anchored: 86\% after one round, and all within three.
\item \emph{On the MOSES set, 1-WL already separates every pair.} The 232,191 MOSES molecules fall into 232,191 different 1-WL classes, so on this set 1-WL, CNOT-\Height{} and both KCZ arms separate exactly the same pairs.
\item \emph{The circuits separate some pairs that 1-WL cannot.} A complete search over graphs with at most 7 atoms found six pairs of different, grammar-accepted molecules that neither plain nor anchored 1-WL can separate. All four circuit arms separate five of them, for example 7-azabicyclo[2.2.1]heptane (\smi{C1CC2CCC1N2}) and dicyclopropylamine (\smi{C1CC1NC1CC1}). The sixth, a strained aza-cage pair, is separated by both KCZ arms but collides under both CNOT arms, because its port-line graphs match under both key schemes. It is the only CNOT-\Height{} collision we found.
\end{enumerate}

\begin{figure}[t]
\centering
\includegraphics[width=0.95\textwidth]{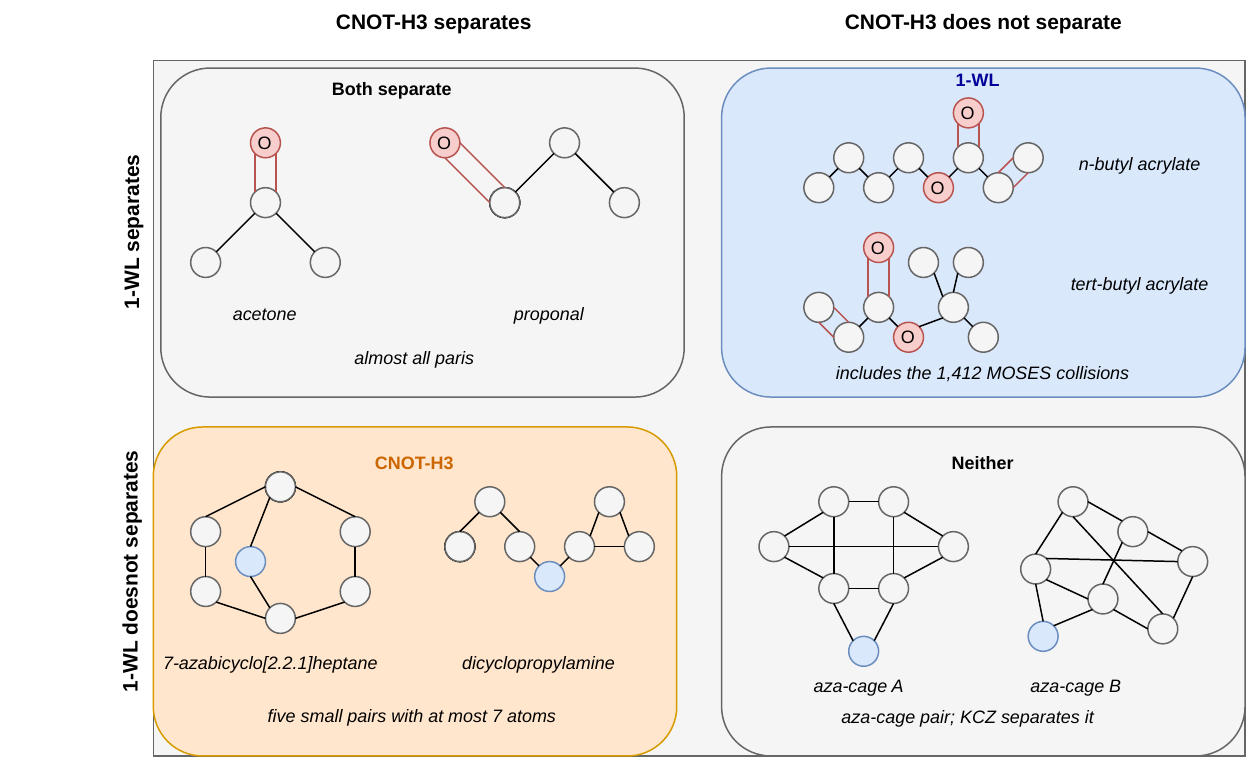}
\caption{Which pairs 1-WL and the tied CNOT-\Hthree{} pipeline can separate, for grammar-accepted molecule pairs in canonical form from Tox21, the generated set, the MOSES set and a complete search over graphs with at most seven atoms. Each panel shows one example pair. \emph{1-WL only}: every certified collision, including the 1,412 MOSES collisions. \emph{CNOT-H3 only}: five small pairs that 1-WL cannot separate but every circuit arm can. \emph{Neither}: an aza-cage pair that KCZ separates. All verdicts are exact.}\label{fig:wl}
\end{figure}

So, with a fixed canonical writing, neither CNOT-\Hthree{} nor 1-WL is stronger than the other: each separates pairs that the other does not (Fig.~\ref{fig:wl}). The tied circuit loses branching information that 1-WL keeps. In the other direction, a sum over the whole network picks up ring structure that colour refinement cannot see, as the bicyclic example shows.

These are certified statements about exact separation on finite sets of molecules. They do not place any arm at a particular level of the WL hierarchy.

\subsection{The collisions do not transfer to standard typed sharing}\label{sec:res-lambeq}

Is this blindness to isomers a feature of compositional circuits in general, or only of our coarse \Hthree{} tying? We tested five sharing schemes in the style of lambeq on the same structures (Table~\ref{tab:lambeq}). None gave a single collision \textbf{[F]}.

\begin{table}[h]
\caption{Collisions under lambeq-style sharing, on all 1,773,786 generated-set pairs, the 55 pairs among the Tox21 collision molecules and the 6 engineered pairs. Every pair received an exact witness.}\label{tab:lambeq}
\begin{tabular*}{\textwidth}{@{\extracolsep\fill}llll}
\toprule
Variant & Word & Sharing & Collisions \\
\midrule
V1-T & IQP, 1 layer & element + full type + slot & 0 / 0 / 0 \\
V2-T & IQP, 2 layers & element + full type + slot & 0 / 0 / 0 \\
V3-T & Sim14 & element + full type + slot & 0 / 0 / 0 \\
V4-T & CNOT ladder & element + full type + slot & 0 / 0 / 0 \\
V1-P & IQP, 1 layer & literal positional names & 0 / 0 / 0 \\
\Hthree{} + arity & CNOT ladder & \Hthree{} refined by word arity & 0 / 0 / 0 \\
\Hthree{} + slot & CNOT ladder & \Hthree{} refined by port slot & 0 / 0 / 0 \\
\botrule
\end{tabular*}
\end{table}

The controls behaved as before: CNOT-\Hthree{} gave the same 8,091 collisions on S1 as in the census, and CNOT-\Height{} gave none. The clearest contrast is within one word. The same CNOT ladder gives 8,091 collisions under \Hthree{} and none under typed sharing (V4-T, 56 symbols). Even splitting the 13 \Hthree{} symbols by arity (21 symbols) or by slot (22 symbols) is enough. So the collisions come from tying angles in a way that ignores arity and position.

Removing collisions and removing dependence on writing are two separate goals. Slot labels follow the port order, and the port order follows the writing, so slot-typed sharing separates the isomers but keeps the CNOT word's dependence on writing. Table~\ref{tab:sharing} puts the main schemes side by side, together with how far apart the repaired outputs really are (Section~\ref{sec:res-budget}).

\begin{table}[t]
\caption{Sharing schemes compared. Symbols are those used on the generated set, and ``exact collisions'' refers to the structures tested for that scheme. $D$ is the median trace distance between normalised output states, and $N_{\mathrm{rec}}$ the median lower bound on preparations when success flags are also recorded, for the seven Tox21 \Hthree-collision pairs over 48 initialisation-law tables (Section~\ref{sec:res-budget}). Theorem~\ref{thm:bound} applies to every row.}\label{tab:sharing}
\footnotesize\begin{tabular*}{\textwidth}{@{\extracolsep\fill}lccccc}
\toprule
Scheme & Symbols & Writing-invariant & Exact collisions & $D$ & $N_{\mathrm{rec}}$ \\
\midrule
CNOT-\Hthree & 13 & no & yes & 0 & $\infty^{a}$ \\
CNOT-\Hthree{} + slot & 22 & no & none & $1.5\times10^{-5}$ & $8.6\times10^{6}$ \\
CNOT-\Hthree{} + arity & 21 & no & none & -- & -- \\
CNOT-\Height & 20 & no & one (cage pair) & $2.0\times10^{-4}$ & $7.1\times10^{5}$ \\
CNOT, typed (V4-T recipe) & 56 & no & none$^{b}$ & $1.0\times10^{-3}$ & $8.5\times10^{5}$ \\
KCZ-\Hthree & 13 & yes (Prop.~\ref{prop:kcz}) & none & $2.2\times10^{-2}$ & $1.2\times10^{6}$ \\
KCZ-\Height & 20 & yes (Prop.~\ref{prop:kcz}) & none & $1.0\times10^{-1}$ & $1.4\times10^{6}$ \\
\botrule
\end{tabular*}
\footnotetext{$^{a}$ For the six strongly equal pairs the success flag and the successful output state are identical, so no measurement on that record separates them; failed-run outcomes were not analysed. The projectively equal ether pair is separated through its success rate (median lower bound 524 preparations). $^{b}$ The certified V4-T parameterisation gives no collision on the generated set (Table~\ref{tab:lambeq}); our implementation of the same recipe, which gives $D$ and $N_{\mathrm{rec}}$, separates exactly all 55 pairs among the Tox21 collision molecules, the $n=16$ pair and all 1,988 MOSES collision pairs of Section~\ref{sec:res-budget}.}
\end{table}

\subsection{Exact separation versus distinguishability at a finite budget}\label{sec:res-budget}

An exact witness shows that two outputs differ at some choice of angles. It does not say whether a shared angle table separates them by a useful amount, or how many runs a real experiment would need. We measured both with independently written code, using exact arithmetic throughout \textbf{[N]}.

\paragraph{Design.} We used the seven Tox21 pairs that collide under \Hthree{} and, as controls, 200 random pairs of accepted Tox21 molecules that are isomers of each other (same formula, different structure). The arms were CNOT words under \Hthree, \Height, \Hthree{} plus slot, and our typed scheme (element, full word type and slot), and KCZ words under \Hthree{} and \Height. Each arm used 48 angle tables drawn from the starting range, with every angle moved to a nearby angle with rational sine and cosine. Every output is then an exact vector of fractions, and the trace distance between the normalised output states is
\begin{equation}
D=\frac{\lvert u_0v_1-u_1v_0\rvert}{\lVert u\rVert\,\lVert v\rVert},
\end{equation}
so checking $D=0$ is exact. Costs were computed from these fractions with 60-digit arithmetic. Cells with an undefined output were left out (16 and 75 of 9,600 control cells for the two KCZ arms).

\paragraph{Costs.} Assuming both molecules are equally likely, and aiming for a 1\% error, we compare:
\begin{itemize}
\item $N_{\mathrm{state}}$, the expected number of runs for a protocol that looks only at the kept output states, preparing until the best possible (Helstrom) measurement \cite{helstrom1976} on those states reaches 1\% error;
\item $N_{\mathrm{rec}}$, a lower bound for any measurement on the full record of each run (whether it succeeded and, if it did, the output qubit). With $p=\Pp^{X}$, $q=\Pp^{Y}$ and kept states $\psi_X,\psi_Y$, the record has fidelity $F_{\mathrm{rec}}=\sqrt{(1-p)(1-q)}+\sqrt{pq}\,\lvert\langle\psi_X|\psi_Y\rangle\rvert$, and the Fuchs--van de Graaf inequality \cite{fuchs1999distinguishability} says the error after $N$ runs is at least $\tfrac12\big(1-\sqrt{1-F_{\mathrm{rec}}^{2N}}\big)$;
\item $N_{\mathrm{rec}}^{+}$, a number of runs that is enough for the same record, from the quantum Chernoff bound \cite{audenaert2007chernoff};
\item $N_{\mathrm{flag}}$ and $N_{\mathrm{flag}}^{+}$, the same two numbers when only the success flag is kept.
\end{itemize}
These bounds are for a fixed number of independent runs at one fixed angle table. Medians are given with bootstrap intervals that resample pairs and tables together.

\paragraph{The repairs barely move the output state.} Under CNOT-\Hthree{} all 336 collision cells have $D=0$ exactly; ordinary floating point would instead return tiny non-zero values. After a repair, the pairs that used to collide have a median $D$ of $2.0\times10^{-4}$ under \Height{} (95\% interval $1.7\times10^{-5}$--$6.2\times10^{-4}$), $1.5\times10^{-5}$ with slot labels and $1.0\times10^{-3}$ with typed sharing, compared with 0.05--0.06 for the controls. A protocol that looks only at the output state would need a median of $4\times10^{10}$ to $6\times10^{13}$ runs, compared with $2$--$3\times10^{7}$ for the controls.

\paragraph{The success rate carries the difference.} When the success flag is recorded as well as the output state, the lower bound drops to a median of $7\times10^{5}$ (\Height), $9\times10^{6}$ (slot) and $8\times10^{5}$ (typed) runs, close to the $5$--$9\times10^{5}$ of the controls. Almost all of this comes from the flag. For every repaired CNOT arm, the median ratio $N_{\mathrm{rec}}/N_{\mathrm{flag}}$ on the collision pairs is 1.00 to two decimals. The number of runs that is provably enough for a flag-only test, $N_{\mathrm{flag}}^{+}$, is a median 2.4 times $N_{\mathrm{rec}}$, and within 2.5 times in 82--91\% of cells (2.4 is simply the gap between our two bounds when the fidelities are equal). For the controls, the flag-only number comes this close in only 25--44\% of cells, which leaves room for the output state to help. So what tells the formerly colliding isomers apart is how often the circuit succeeds. A readout trained on the normalised output qubit, like ours and like standard lambeq-style classifiers, never sees this.

\paragraph{A larger test on MOSES.} Seven pairs is a small sample, so we repeated the analysis on MOSES with our independent code \textbf{[N]}. This code accepts a somewhat wider range of molecules than the main compiler (531,517 molecules rather than 232,191). Searching for matching port-line graphs, and confirming each match exactly, gave 1,997 strongly equal CNOT-\Hthree{} pairs. Nine of them are tautomers whose heavy-atom graphs are identical, so every repair keeps them equal, and we set them aside. Each of the remaining 1,988 pairs is separated exactly by \Height, by slot labels and by typed sharing. Over 48 tables per pair, and against 300 random MOSES control pairs of isomers, the Tox21 picture holds and becomes sharper. The repaired output states differ by a median $D$ of $2.3\times10^{-8}$ (\Height), $1.4\times10^{-9}$ (slot) and $5.5\times10^{-8}$ (typed), compared with 0.035--0.044 for the controls, so a state-only protocol would need $10^{23}$--$10^{26}$ runs instead of about $10^{12}$. The median ratio $N_{\mathrm{rec}}/N_{\mathrm{flag}}$ is 1.00 in every repaired arm, and the flag-only number of runs is within a factor 2.5 of the full-record lower bound in 95--96\% of cells, compared with 46--54\% for the controls (Table~\ref{tab:moses}). These larger molecules also succeed so rarely that every pair, collision or control, needs about $10^{10}$--$10^{12}$ runs even under the record bound.

\begin{table}[h]
\caption{Number of runs needed on MOSES (independent code): 1,988 CNOT-\Hthree{} collision pairs and 300 control pairs of isomers, 48 tables per pair, exact arithmetic. ``Flag $\le2.5\times$'' is the share of cells in which the flag-only number of runs that is enough, $N_{\mathrm{flag}}^{+}$, is within a factor 2.5 of the full-record lower bound $N_{\mathrm{rec}}$. Under CNOT-\Hthree{} every collision cell has $D=0$.}\label{tab:moses}
\footnotesize\begin{tabular*}{\textwidth}{@{\extracolsep\fill}lcccccccc}
\toprule
& \multicolumn{2}{c}{Median $D$} & \multicolumn{2}{c}{Median $N_{\mathrm{state}}$} & \multicolumn{2}{c}{Median $N_{\mathrm{rec}}$} & \multicolumn{2}{c}{Flag $\le2.5\times$} \\
Repair & collision & control & collision & control & collision & control & collision & control \\
\midrule
\Height & $2.3\times10^{-8}$ & 0.039 & $1.2\times10^{24}$ & $2.4\times10^{12}$ & $4.0\times10^{10}$ & $2.2\times10^{10}$ & 96\% & 54\% \\
slot & $1.4\times10^{-9}$ & 0.035 & $3.7\times10^{26}$ & $3.3\times10^{12}$ & $9.5\times10^{11}$ & $5.2\times10^{10}$ & 95\% & 49\% \\
typed & $5.5\times10^{-8}$ & 0.044 & $2.5\times10^{23}$ & $2.1\times10^{12}$ & $1.0\times10^{11}$ & $3.0\times10^{10}$ & 96\% & 51\% \\
\botrule
\end{tabular*}
\end{table}

\paragraph{Which compiler?} The independent code follows the rules of Section~\ref{sec:construction}, but in this run it only checked that aromatic rings were separate and fully aromatic, which is why it accepts more molecules. Limited to the scope of Section~\ref{sec:scope}, it accepts 250,670 MOSES molecules, against 232,191 for the main compiler, and finds 1,456 port-line collision pairs covering 1.14\% of them, against 1,412 pairs and 1.19\%. The pair list of the main census was not used in this re-analysis, so the two lists have not been compared pair by pair, and we have not tracked down the remaining 8\% difference in acceptance. On the 1,453 restricted pairs that are not tautomers, the results do not change: median $D$ of $1.5\times10^{-9}$ to $5.4\times10^{-8}$, median ratio $N_{\mathrm{rec}}/N_{\mathrm{flag}}$ of 1.00, and the flag-only number of runs within a factor 2.5 in 96--97\% of cells.

\paragraph{After training.} Everything above uses untrained tables. To see whether training changes this, we trained the three repaired CNOT arms on Tox21 NR-ER with our independent code, using the recipe of Section~\ref{sec:data} cut down to 300 Adam steps, five folds and one starting table per fold. We then measured the same pairs at the starting and at the trained table of each fold \textbf{[N]}. Every port in these pairs has a key that also appears in the training molecules, so every angle they use was trained. Table~\ref{tab:trained} gives the result. Training pushes ordinary isomers apart: the median $D$ of the controls rises from 0.04--0.06 to 0.08--0.14. It does not do this for the formerly colliding pairs, whose median $D$ stays between $4\times10^{-5}$ and $6\times10^{-4}$ and even falls slightly in every arm. The success flag still carries all of their difference: the median ratio $N_{\mathrm{rec}}/N_{\mathrm{flag}}$ is 1.00, and the flag-only number of runs is within a factor 2.5 of the full-record lower bound in 86--97\% of cells. On this task, then, training does not create the separation in the output state that the repairs fail to provide.

\begin{table}[h]
\caption{Before and after training (independent code; 7 Tox21 collision pairs and 200 controls; five folds, one starting table each, 300 Adam steps). ``Flag $\le2.5\times$'' as in Table~\ref{tab:moses}.}\label{tab:trained}
\footnotesize\begin{tabular*}{\textwidth}{@{\extracolsep\fill}llcccc}
\toprule
& & \multicolumn{2}{c}{Median $D$} & & \\
Repair & Table & collision & control & Median $N_{\mathrm{rec}}/N_{\mathrm{flag}}$ (collision) & Flag $\le2.5\times$ (collision) \\
\midrule
\Height & initial & $4.8\times10^{-4}$ & 0.063 & 1.00 & 91\% \\
 & trained & $1.8\times10^{-4}$ & 0.099 & 1.00 & 97\% \\
\Hthree{} + slot & initial & $5.7\times10^{-5}$ & 0.041 & 1.00 & 74\% \\
 & trained & $4.1\times10^{-5}$ & 0.077 & 1.00 & 94\% \\
typed & initial & $9.7\times10^{-4}$ & 0.047 & 1.00 & 86\% \\
 & trained & $6.0\times10^{-4}$ & 0.140 & 1.00 & 86\% \\
\botrule
\end{tabular*}
\end{table}

\paragraph{The symmetric word.} KCZ separates the output states much better (median $D$ of 0.022 and 0.10 on the collision pairs, 0.29 and 0.45 on the controls) but succeeds less often (median $\min\Pp$ of $2\times10^{-6}$ on the \Height{} collision pairs). So it still needs $2$--$3\times10^{8}$ runs with the state alone, and about $10^{6}$ under the record bound. Under CNOT-\Hthree, nothing in the record separates the six strongly equal pairs, while the ether pair is separated by its success rate alone (median lower bound 524 runs).

The controls were drawn from all accepted Tox21 molecules, not only from the development set of Table~\ref{tab:collisions}. One control pair, 2-methylpropane-1,3-diol and butane-1,3-diol, has $D=0$ exactly at all 48 tables. Its port-line graphs match, with an explicit mapping, so it is one more CNOT-\Hthree{} collision certified by Corollary~\ref{cor:portline}, this time among Tox21 molecules outside the development set. In addition, 19 of the 200 controls fall below $D=10^{-7}$ at some tables under CNOT-\Hthree; these are mostly long-chain isomers that differ far from the anchor.

\subsection{A cup-free baseline}\label{sec:res-cupfree}

Designs that place ordinary (unitary) gates on bonds need no postselection \cite{pegg2026topology}. To see what this changes, we ran a small model of our own in this style on the same pairs \textbf{[N]}. Each heavy atom gets one qubit, prepared by $R_y$ with an angle keyed by its label (or by its label and number of neighbours). Then come $L$ layers. Each layer applies a controlled-phase gate on every bond, keyed by bond type, followed by an $R_y$ on every atom. The readout is the state of the anchor qubit, with the anchor chosen as before. The circuit depends only on the labelled graph and the anchor, so by design it does not depend on the writing, and every run gives one copy of the output. We used 48 tables per arm, the seven collision pairs and the 178 control pairs with at most 14 heavy atoms, and computed $D$ and $N_{\mathrm{rec}}$ (now for the anchor state alone) from exact state vectors in double precision.

\begin{table}[h]
\caption{Cup-free baseline on the seven Tox21 \Hthree-collision pairs and 178 control pairs, 48 tables per arm. The anchored readout is the anchor qubit; the pooled readout is the average of the states of all atom qubits. ``Tied'' counts pairs whose readout states are identical at every table. Medians are over all cells; $\infty$ means that more than half the cells are tied.}\label{tab:cupfree}
\footnotesize\begin{tabular*}{\textwidth}{@{\extracolsep\fill}llccccccc}
\toprule
& & & \multicolumn{2}{c}{Tied pairs} & \multicolumn{2}{c}{Median $D$} & \multicolumn{2}{c}{Median $N_{\mathrm{rec}}$} \\
Readout & Keys & $L$ & collision & control & collision & control & collision & control \\
\midrule
anchored & label & 2 & 7/7 & 29/178 & 0 & 0.17 & $\infty$ & 83 \\
anchored & label & 6 & 1/7 & 4/178 & $7.6\times10^{-3}$ & 0.22 & $4.9\times10^{4}$ & 62 \\
anchored & label + degree & 2 & 5/7 & 19/178 & 0 & 0.25 & $\infty$ & 41 \\
anchored & label + degree & 6 & 1/7 & 3/178 & $2.2\times10^{-2}$ & 0.24 & $5.5\times10^{3}$ & 51 \\
\midrule
pooled & label & 2 & 0/7 & 0/178 & 0.031 & 0.045 & $3.0\times10^{3}$ & $1.4\times10^{3}$ \\
pooled & label & 6 & 0/7 & 0/178 & 0.041 & 0.056 & $1.9\times10^{3}$ & 993 \\
pooled & label + degree & 2 & 0/7 & 0/178 & 0.064 & 0.082 & 784 & 443 \\
pooled & label + degree & 6 & 0/7 & 0/178 & 0.062 & 0.072 & 807 & 605 \\
\botrule
\end{tabular*}
\end{table}

Without postselection, control pairs need a few dozen runs instead of the $5$--$9\times10^{5}$ of the cup encoders (Table~\ref{tab:cupfree}). This design has its own blind spot, and it is about distance. A qubit can affect the anchor only through a chain of gates, and each layer lengthens the chain by one bond. So the anchor's state depends only on the labelled atoms within $L$ bonds of the anchor and the bonds between them, except bonds that join two atoms both exactly $L$ bonds away \textbf{[M]}. Every tie in the run fits this rule. With two layers, all seven collision pairs are tied under label keys; with six, isooctyl and n-octyl acrylate still are, because they differ only at the far end of the chain. Where the baseline does separate the formerly colliding isomers, it separates them much less than the controls, though with $10^{3}$--$10^{4}$ runs rather than about $10^{6}$. In the cup encoders, postselection links the anchor to every atom, and \Height{} separates all seven pairs.

This limited reach comes from reading only the anchor, not from the circuit itself. If we instead read the average of all single-atom states, which is the same as measuring one atom qubit picked at random, no anchor is needed and the result still does not depend on the writing. With this pooled readout no pair is tied in any cell, already at two layers, and the formerly colliding isomers are separated almost as well as the controls (median $D$ of 0.03--0.06 against 0.045--0.08). The price is that averaging waters down every difference, so the controls now need about $10^{3}$ runs rather than a few dozen. For this baseline, then, the success-rate effect of Section~\ref{sec:res-budget} does not arise, and once the readout looks at every atom the formerly colliding isomers are no harder to separate than ordinary isomers. Whether a similar readout over many atoms would help the cup encoders, whose single output is fixed by the grammar, is an open question.

\subsection{Norms and postselection in practice}\label{sec:res-norm}

The trained CNOT circuits use a median of 25 qubits (at most 91), all but one of them postselected. Under \Height, the median $\Pp$ per fold among rows with a defined output ranged from $2.9\times10^{-7}$ to $9.3\times10^{-6}$. Fitting $\log_2\Pp$ against the number of bonds gives a slope of $-1.50$ per bond, of which $-1$ comes directly from $\Pp=2^{-n}\Pt$ \textbf{[N]}. On 2,304 accepted Tox21 molecules, 23,390 evaluations with CNOT, KCZ and random normalised words gave a largest $\Pt$ of 0.965, so the bound of Theorem~\ref{thm:bound} always held; molecules with rings stayed well below it (data in the code package).

\begin{figure}[t]
\centering
\includegraphics[width=\textwidth]{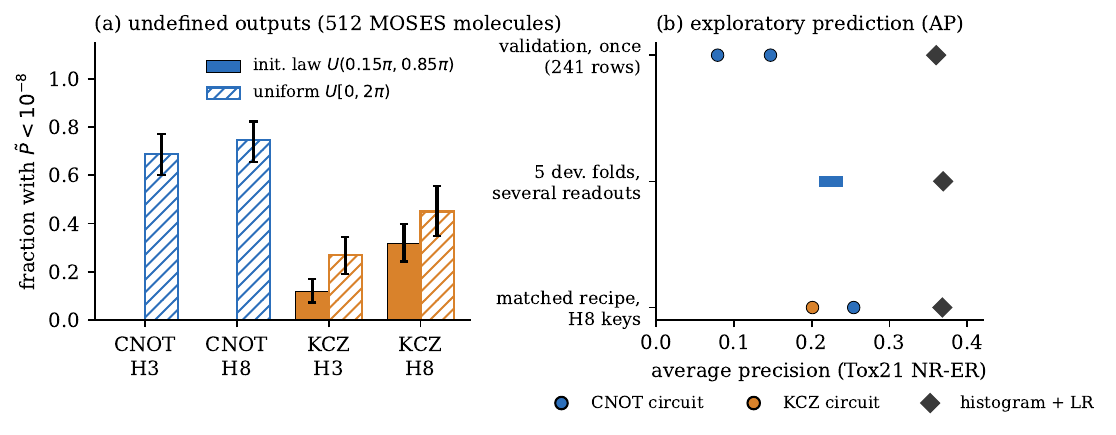}
\caption{(a) Share of 512 MOSES molecules with an undefined output ($\Pt<10^{-8}$), over 32 tables per setting, with bootstrap 95\% intervals. With angles from the starting range, KCZ loses norm far more often than CNOT, because its terms cancel (Theorem~\ref{thm:ising}). (b) Exploratory prediction on Tox21 NR-ER: every circuit scores below a logistic regression on a simple count of (label, degree) pairs, which ignores order. The middle-row bar shows the range over readouts; these are single training runs, so there are no intervals.}\label{fig:normpred}
\end{figure}

Symmetry costs norm. Over the 1,299 Tox21 molecules, KCZ raises the number of two-qubit gates inside words from 17,936 to 22,429. At the shared starting table it also lowers the median $\log_{10}\Pt$ under \Height{} from $-0.73$ to $-3.15$. With angles from the starting range, 12\% of KCZ-\Hthree{} outputs and 32\% of KCZ-\Height{} outputs are undefined, against almost none for CNOT (Fig.~\ref{fig:normpred}a). The reason is the sign pattern of Theorem~\ref{thm:ising}. In this range every CNOT amplitude is a sum of positive terms and cannot cancel \textbf{[M]}, while every undefined KCZ cell shows strong cancellation; the same networks with all terms made positive are undefined in only 0.1\% of those cells \textbf{[N]}. With angles drawn uniformly the order reverses (69--75\% undefined for CNOT, 27--45\% for KCZ).

\subsection{Prediction (exploratory)}\label{sec:res-prediction}

On Tox21 NR-ER, every circuit model we trained scores well below a logistic regression on a simple count (histogram) of (atom label, number of heavy-atom neighbours) pairs, which ignores order \textbf{[X]}. Fig.~\ref{fig:normpred}b summarises the comparison and Table~\ref{tab:prediction} gives the numbers. They are average precision (AP) on the reused development folds unless stated otherwise; a model with no skill scores about 0.06--0.14, depending on the split.

\begin{table}[h]
\caption{Exploratory predictive comparison on Tox21 NR-ER (average precision).}\label{tab:prediction}
\begin{tabular*}{\textwidth}{@{\extracolsep\fill}p{5.6cm}p{3.4cm}p{2.4cm}}
\toprule
Comparison & Circuit & Histogram + LR \\
\midrule
Scaffold-held-out validation, used once (241 rows, 15 positive) & 0.079 (\Hthree), 0.147 (\Height) & 0.360 \\
Five development folds, several readouts & 0.21--0.24 (AUC $\approx$ 0.60) & 0.369 (AUC 0.742) \\
Adding the low-norm hinge (\Height) & undefined rows 76 $\to$ 17; AP $+0.040$ in 4 of 5 folds & ahead in every fold \\
CNOT vs KCZ, \Height, matched recipe & CNOT 0.254, KCZ 0.201 & 0.368 \\
\botrule
\end{tabular*}
\end{table}

The histogram beat the KCZ model in 9 of 10 combinations of key scheme and fold. CNOT and KCZ differ not only in invariance but also in their gates, starting norm and rate of undefined outputs, so the difference between them cannot be put down to invariance alone. Adding fixed circuit features to the histogram did not help in a consistent way.

Where do the trained circuits ``look''? In 25 trained circuits, the output was sensitive mainly to small changes in gates near the anchor: under \Height{}, the median share of sensitivity from gates more than two bonds away was $7.0\times10^{-6}$ \textbf{[N]}. A histogram limited to the anchor's neighbourhood scores 0.148 AP lower than the whole-molecule histogram in every fold. This fits, but does not prove, the idea that the circuits do poorly because they only look locally.

These results come from five development folds that were reused throughout the project, one starting table per fold and a single use of the validation split, and they do not include V4-T sharing. On the validation split the circuits are close to chance: CNOT-\Hthree{} reaches an AP of 0.079, against a no-skill level of about 0.062. We use them only as a comparison with the histogram baseline. A fragment-level version of the grammar, which we built but did not train, is described in Supplementary Note~S3.

%% =====================================================================
\section{Discussion}\label{sec:discussion}

\paragraph{What the grammar contributes.} The pregroup types make the wiring match the molecular graph and decide which ports share angles. The types are read from the graph, but they do not enforce valence, and they do not choose the gates, the anchor or the readout. On their own, then, they do not make the model depend only on the molecule. Unlike in the MOF study, the composition here really is chemical, with one contraction per bond; but that alone did not beat a simple count that ignores order.

\paragraph{Can the three grammars be aligned?} Partly. Cups along bonds line up the compositional grammar with how the atoms are connected. Checking chemical validity would need word types that carry valence, which we did not build. And the grammar of the written string comes back in two ways: through the order of the ports, and because acceptance is decided on the string. A symmetric word removes the first of these, at a cost in norm and gates, but not the second.

\paragraph{What is new, and what is not.} The tools we use are standard: spin-model partition functions, correspondences between trees, quadratic forms over GF(2), the fact that matrix norms multiply at most, and the rules for moving Pauli corrections through Clifford circuits. Invariance from symmetric gates that can be applied in any order is already known \cite{mernyei2022eqgc}, and order-free models beating DisCoCat on chemical data had already been reported \cite{kang2025mof}. What this paper adds is an exact account of one clearly defined class of models. The port-line graph explains all but one of the collisions we certified on real molecule sets. The shape of a side chain fully determines the output up to 15 atoms but not from 16, and the smallest counterexample is produced by the compiler itself. Repairs that remove the collisions leave the output states almost identical, so the success rate becomes the main signal, and a standard readout throws it away. The tied circuit and 1-WL each separate pairs the other cannot. A cup-free baseline avoids the cost of postselection; its anchored readout can only see a limited distance, and reading all atoms removes this limit. The general lesson is not that compositional encoders must have collisions, since typed sharing avoids them. It is that what an encoder keeps, and what it costs to read out, can be checked exactly before any training.

\paragraph{Why the split between flag and state matters.} The MOSES numbers cut both ways, and we say so plainly. At drug-like sizes, every pair we measured, controls included, needs about $10^{10}$--$10^{12}$ runs through a cup encoder even when the success flag is used. In practice, then, none of these molecules can be told apart this way. The split between flag and state still matters, for three reasons. First, it shows where the information added by a repair actually goes: into the success rate, which standard lambeq-style classifiers throw away because they read only the normalised output qubit. A model can therefore be exactly one-to-one on paper and still give a trained readout almost nothing to work with. Second, it gives a test that any repaired design should pass before training: measure the trace distance of the output states and the difference in success rate separately, and check which of the two the readout can use. Third, it is an argument for designs without postselection. The pooled cup-free baseline separated the same Tox21 pairs with about $10^{3}$ runs, and its formerly colliding pairs were almost as easy to separate as ordinary isomers.

\paragraph{Practical checks.} Our results suggest five checks for compositional molecular circuits:
\begin{enumerate}
\item Compare cup endpoints with the RDKit bond list for every string.
\item Do not share angles by element and bond type alone. Splitting by arity or slot costs only about eight extra symbols. After such a repair, measure the trace distance and the difference in success rate separately, and say which one the readout can use.
\item Test directly whether the output changes when the molecule is written differently, with exact arithmetic where possible.
\item Report $\Pp$, not only $\Pt$, and expect at least $2^{n}$ runs for each kept run. If a cup-free design is used instead, check how far its readout can see, or read all atoms.
\item Compare with a simple count that ignores order before crediting any result to composition.
\end{enumerate}

\paragraph{Scope and limitations.} All results are about the compiler, word types, sharing schemes and molecule sets described here. The chemistry covered is limited: no charges, stereochemistry, isotopes, most elements, or fused and multi-ring aromatic systems, and acceptance depends on the written string. The MOSES census uses canonical writings only. Prediction was tested on one task with reused folds, so it is exploratory. The finite-budget and cup-free analyses use untrained tables, apart from one short training run with our independent code (300 steps, one starting table per fold), and they assume ideal measurement statistics. The MOSES part of the finite-budget study uses our independent code, whose scope differs from that of the main compiler, and we ran no experiments on hardware or with noise. The 1-WL comparison rests on six pairs of at most seven atoms. Two questions remain open: whether longer training, other tasks or other starting tables would widen the separations that repairs create, and whether failed Bell outcomes carry further useful information.

%% =====================================================================
\section{Conclusion}\label{sec:conclusion}

Giving each atom a type built from its bonds, and closing one cup per bond, produces quantum circuits whose wiring is exactly the molecular graph. It does not produce a model whose output depends only on the molecule. When angles are shared by element and bond type, the output is a partition function on a port-line graph. This single object explains all but one of the collisions we certified on real molecules, including 1.19\% of 232,191 MOSES molecules; for constructed alkyl chains it stops being the full story at 16 side-chain atoms. Typed sharing removes the collisions, but the repaired isomers still have almost identical output states, before and after training, and differ mainly in how often postselection succeeds. With a fixed writing, the tied circuit and 1-WL each separate pairs the other cannot. Three simpler results complete the picture. A word that depends on port order lets the SMILES writing leak into trained outputs; a symmetric word stops this, at a cost in norm; and every kept run costs at least $2^{n}$ runs, which a cup-free design avoids, although its anchored readout then sees only a limited distance unless all atoms are read. All of these can be checked exactly before training, and we offer them as a checklist for compositional molecular encoders.

%% =====================================================================
\backmatter

\bmhead{Supplementary information}
Supplementary File S2 contains the full proofs, with every assumption checked case by case. Supplementary Note S3 describes how the grammar was developed, the MOSES census pipeline, the analysis protocols and the corrections made during the project, and the fragment-level version. Certificates and angle tables accompany the submission.

\bmhead{Acknowledgements}
To be supplied by the authors.

\section*{Declarations}

\begin{itemize}
\item \textbf{Funding:} To be supplied.
\item \textbf{Conflict of interest/Competing interests:} To be supplied.
\item \textbf{Ethics approval and consent to participate:} Not applicable.
\item \textbf{Consent for publication:} Not applicable.
\item \textbf{Data availability:} Tox21 \cite{wu2018moleculenet} and MOSES \cite{polykovskiy2020moses} are public. Labels are not redistributed.
\item \textbf{Materials availability:} Not applicable.
\item \textbf{Code availability:} The independent re-implementation scripts, the saved results of Sections~\ref{sec:res-budget} and \ref{sec:res-cupfree} and a quick-start script that runs the diagnostics of this paper (graph faithfulness, writing invariance, exact collision verdicts, output-state distances and preparation budgets) on any list of SMILES strings are provided as supplementary material. Certificates and evidence packages of the main pipeline will be released at a public repository with an archival DOI (to be supplied).
\item \textbf{Author contribution:} To be supplied.
\item \textbf{Use of generative AI:} See Section~\ref{sec:data}, ``Use of AI tools''.
\end{itemize}

%% =====================================================================
\begin{appendices}

\section{Tox21 collision molecules}\label{app:tox21}

\begin{table}[h]
\caption{Molecules involved in the seven certified CNOT-\Hthree{} collisions on the Tox21 development set.}\label{tab:tox21coll}
\begin{tabular*}{\textwidth}{@{\extracolsep\fill}lll}
\toprule
Molecule & Canonical SMILES & Collides with \\
\midrule
Dibutyl ether & \smi{CCCCOCCCC} & Tetrahydrofuran (up to scale) \\
Tetrahydrofuran & \smi{C1CCOC1} & Dibutyl ether \\
2,4-Diaminotoluene & \smi{Cc1ccc(N)cc1N} & 2,6-Diaminotoluene \\
2,6-Diaminotoluene & \smi{Cc1c(N)cccc1N} & 2,4-Diaminotoluene \\
tert-Butyl acrylate & \smi{C=CC(=O)OC(C)(C)C} & iso- and n-butyl acrylate \\
Isobutyl acrylate & \smi{C=CC(=O)OCC(C)C} & tert- and n-butyl acrylate \\
n-Butyl acrylate & \smi{C=CC(=O)OCCCC} & tert- and isobutyl acrylate \\
Isooctyl acrylate & \smi{C=CC(=O)OCCCCCC(C)C} & n-Octyl acrylate \\
n-Octyl acrylate & \smi{C=CC(=O)OCCCCCCCC} & Isooctyl acrylate \\
2,6-Toluene diisocyanate & \smi{Cc1c(N=C=O)cccc1N=C=O} & 2,4-Toluene diisocyanate \\
2,4-Toluene diisocyanate & \smi{Cc1ccc(N=C=O)cc1N=C=O} & 2,6-Toluene diisocyanate \\
\botrule
\end{tabular*}
\end{table}

\clearpage
\section{Claims and their evidence}\label{app:claims}

\begin{table}[h]
\caption{Each main claim, its evidence class and the evidence behind it.}\label{tab:claims}
\begin{tabular*}{\textwidth}{@{\extracolsep\fill}p{5.3cm}lp{5.2cm}}
\toprule
Claim & Class & Evidence \\
\midrule
Cups equal RDKit bonds for every accepted string & F & Routing guard on 12,614 + 2,524 strings and 232,191 identities \\
Trained CNOT outputs depend on the writing & N & Five-fold traversal analysis (TVF 0.224) \\
KCZ outputs are writing-invariant under A1--A5 & M, F & Proof; 2,852 certified equalities; 50,456 numerical checks \\
CNOT outputs are port-line partition functions & M & Proof; independent spin-sum checks \\
Shape factorisation; complete to $n=15$, incomplete from $n=16$ & M, F & Proofs; exhaustive check; four-way certified counterexample \\
1,412 MOSES pairs collide, all others separated & F & Exact witnesses; two certification routes; independent verifier \\
No collisions under five lambeq-style variants & F & Exact witnesses on every tested pair \\
KCZ norm loss is sign cancellation & M, N & Signed Ising theorem; 512-molecule study \\
$\Pp\le 2^{-n}$ under (E1)--(E3) & M, N & Proof; 23,390 independent evaluations \\
Repairs leave output states nearly identical, also after a short training run; the success rate carries the difference & N & Tox21: 7 collision pairs, 200 controls, 6 arms; MOSES: 1,988 collision pairs, 300 controls, 4 arms; 48 tables, exact arithmetic \\
Cup-free baseline: no postselection; anchored readout tied by a light cone, pooled readout separates every tested pair & M, N & Light-cone argument; 7 collision pairs, 178 controls, 8 arms, 48 tables \\
Circuits trail an order-free histogram & X & Development folds; one validation use \\
CNOT-\Hthree{} and 1-WL separations are incomparable & F & Exact 1-WL on 10,092 records and 232,191 identities; verified certificates for six pairs \\
\botrule
\end{tabular*}
\end{table}

\FloatBarrier
\end{appendices}

% References included directly (generated by BibTeX from references.bib) so arXiv does not need to run BibTeX

\end{document}